\documentclass{aa}
\usepackage{times}
\usepackage{epsfig}
\begin{document}

   \thesaurus{04  
              {03.20.1 
               10.19.1 
               10.15.1 
               10.11.1 
               10.19.3 }}

   \title{The distribution of nearby stars in phase space mapped by Hipparcos\thanks{Based on data from the Hipparcos astrometry satellite}}

   \subtitle{II. Inhomogeneities among A-F type stars}

   \author{E. Chereul\inst{1}, M.~Cr\'ez\'e\inst{1,2} \and O. Bienaym\'e\inst{1}}

    \institute{Centre de Donn\'ees de Strasbourg, Observatoire Astronomique de Strasbourg,
11, rue de l'Universit\'e, F-67000 Strasbourg, France
    \and  I.U.P de Vannes, Tohannic, rue Yves Mainguy,  F-56 000 Vannes, France}

    \offprints{chereul@cdsxb6.u-strasbg.fr}

    \date{Received / accepted}
   \authorrunning{E. Chereul et al}
   \titlerunning{The distribution of nearby stars in phase space mapped by Hipparcos. II}
   \maketitle
  \begin{abstract}
A volume limited and absolute magnitude limited sample of A-F type dwarfs within 125 parsecs of the Sun
is searched for inhomogeneities in the density-velocity space, expecting signatures of the cluster evaporation, 
phase mixing and possible disc heating mechanisms.
A 3-D wavelet analysis is used to extract inhomogeneities, both in the density and velocity distributions. 
Thus, a real picture of the phase space is produced. Not only are some clusters and streams detected, 
but the fraction of clumped stars can be measured. 
By estimating individual stellar ages one can relate the streams and clusters to the state of the 
interstellar medium (ISM) at star formation time and provide a quantitative view of cluster evaporation and 
stream mixing. As a result, we propose a coherent interpretation  of moving groups or so-called {\it superclusters} 
and derive some quantitative evolutionary tracers which we expect to serve in the 
understanding of the large scale evolution of the galactic disc.\\
The sample is relatively well mixed in the position space since less than 7 per cent of the stars are proper motion 
confirmed cluster members. We also detect star evaporation out of the Hyades open cluster.\\
Only two components of the velocity vectors are provided by Hipparcos measurements. 
Then, the 3D velocity field is reconstructed from a statistical convergent point method. The wavelet analysis 
exhibits strong velocity structuring at typical scales of velocity dispersion 
$\overline{\sigma}_{stream}\sim$ 6.3, 3.8 and 2.4 $km\cdot s^{-1}$. 
The majority of large scale velocity structures ($\overline{\sigma}_{stream}\sim$ 6.3 $km\cdot s^{-1}$) are 
Eggen's {\em superclusters}. As illustrated by the Pleiades {\em supercluster} these structures are all 
characterized by a large age range which reflects the overall sample age distribution. These 
large velocity dispersion structures represent 63$\%$ of the sample. 
This percentage drops to 46$\%$ if we subtract the velocity background expected by a smooth velocity ellipsoid in 
each structure. Smaller scales ($\overline{\sigma}_{stream}\sim$ 3.8 and 2.4 $km\cdot s^{-1}$) reveal 
that {\em superclusters} are always substructured by 2 or more streams which generally exhibit a coherent age 
distribution. At these scales, the contribution of background stars is negligible and percentages of stars in streams are 38$\%$ 
and 18$\%$ respectively. The detailed analysis of the phase space structures provides a scenario of kinematical 
evolution in the solar neighbourhood: star formation in the galactic disc occurs in large bursts (possibly 
subdivided into smaller bursts) separated 
by quiescent periods. The velocity space is gradually populated by these star formation bursts 
which preferentially fill the center of the velocity ellipsoid. Stars form in groups reflecting the clumpy 
structure of the ISM: about 75$\%$ of recently formed stars belong to streams whose
internal velocity dispersions do not exceed 4 $km\cdot s^{-1}$. Most of them dissolve rapidly. A fraction of the 
initial groups are 
gravitationally bound and form open clusters. Open clusters sustain a longer term streaming with quite similar velocity 
by an evaporation process due to internal processes or encounters with permanent or transient large mass concentrations. 
These streams are detected with $\overline{\sigma}_{stream}\sim$ 2.3 and 3.8 $km\cdot s^{-1}$ and have a 
coherent age content. This process explains the survival of streams up to $10^{9}$ yr. The existence of streams as 
old as 2 Gyr seems to require other physical mechanisms.
The typical scale of so-called Eggen's {\em superclusters} ($\overline{\sigma}_{stream}\sim$ 6.3 $km\cdot s^{-1}$) 
does not seem to correspond to any physical entity. The picture they form, their frequency and their divisions at 
smaller scales are well compatible with their creation by chance coincidence of physically homogeneous smaller 
scale structures  ($\overline{\sigma}_{stream}\sim$ 3.8 or 2.4 $km\cdot s^{-1}$). 
   \end{abstract}
%
\section{Introduction}
\subsection{A tentative observation of the stellar gas kinetics}
Providing a complete probe of the kinematics of early type stars within a well defined local 
sphere, Hipparcos data \cite{Hip97} offer the first opportunity to get an unbiased look into the 
stellar kinematics at small scales. In paper I (\cite{Creze98}), we used the average 
density trend and vertical velocity dispersion to map the potential well across the galactic plane. 
Beyond this 0-order analysis, we now try to get a picture of the kinematic mixture, its 
inhomogeneities  at small scales and their dating. This paper explains the methods used and gives 
the general results obtained with our sample. Details both in the methodology and the results can always be 
found in \cite{Chereul98} (hereafter Paper III). A fourth paper will be dedicated to the physical 
interpretation of evolutionary aspects.\\
Wavelet analysis is extensively used to extract deviations from smooth homogeneity, both in 
density (clustering) and velocity distribution (streaming). A 3-D wavelet analysis tool is first 
developed and calibrated to recognize physical inhomogeneities among random fluctuations. It 
is applied separately in the position space (Section \ref{sec:density}) and in the velocity space 
(Section \ref{sec:velocity}). Once significant features are isolated (whether in density or in velocity), 
feature members can be identified and their behaviour can be traced in the complementary space 
(velocity or density). Thus, a real picture of the phase space is produced. Not only are some clusters 
and streams detected, but the fraction of stars involved in clumpiness can be measured. 
Then, estimated stellar ages help connecting streams and clusters to the state of the ISM at star formation 
time and providing a quantitative view of cluster melting and stream mixing at work.\\
Only once this picture has been established on a prior-less basis we come back to previously 
known observational facts such as clusters and associations, moving groups or so-called 
{\it superclusters}. As a result, we propose a coherent  interpretation  of this phenomenology  
and derive some quantitative evolutionary tracers which we expect to serve in the 
understanding of the large scale evolution of the galactic disc.\\
A sine qua non condition for this analysis to make sense is the completeness of data within well 
defined limits. In so far as positions, proper motions, distances, magnitudes and colours are concerned, 
the Hipparcos mission did care. Things are unfortunately not so simple concerning 
radial velocities and ages. More than half radial velocities are missing and the situation regarding 
Str\"omgren photometry data for age estimation is not much better. So we developed palliative 
methods which are calibrated and tested on available true data to circumvent the completeness 
failure. The palliative for radial velocities is based on an original combination of the classical 
convergent point method with the wavelet analysis, it is presented in Paper III, Section 5.1. The palliative for 
ages (Section \ref{sec:agedet}) is an empirical relationship between age and an absolute-magnitude/colour 
index. It has only statistical significance in a very limited range of the HR diagram.
\subsection{Review of known facts}
While clumpy at the time of star formation, the distribution of stars in phase space gradually
evolves towards a state of smoother homogeneity. The precise mechanism of this mixing
is not elucidated: there are reports of moving groups (also named {\em superclusters} by Eggen) 
with large velocity dispersions (6-8 $km\cdot s^{-1}$) surviving after many galactic rotations 
(Eggen 1991, 1992abc) although not obviously bound by internal gravitation. Considering the very 
short phase mixing time (a few 10$^{8}$ yr, \cite{Hen64}) such streams, exhibiting also large age 
ranges, cannot be explained by the fact that stars originate in a same cell of the phase space.\\
The study of these moving groups is extensive. A large part of the recent 
work deals with their precise identification among early-type (O through F spectral types) 
main sequence stars (Gomez et al, 1990, Chen et al, 1997, Figueras et al, 1997, and Chereul et al, 1997) 
which permit both to probe them far from the Sun and to have an individual age estimation using 
Str\"omgren photometry (\cite{Figue91} and Asiain et al, 1997). Nevertheless, their age content which 
sometimes spreads over several 
hundred million years, as noticed in Eggen's work, is still a puzzle. Two main explanations have been suggested.
On one hand, a {\it supercluster} is formed from several star formation bursts occurring at different times in a 
common molecular cloud (Weidemann et al, 1992) which experiences perturbations such as large 
scale spiral shocks (Comer\'on et al, 1997). These bursts can be gravitationally bound and maintain streams 
by cluster evaporation process during their lifetime. Dynamics of star clusters and giant molecular clouds 
(hereafter GMC) in the galactic disc are different. Thus, different star formation episodes in a same GMC 
separated by several galactic rotation periods will hardly result in giving a {\em supercluster} velocity structure. 
On other hand, a{\it supercluster} is a chance superposition of several cluster evaporations or remnants (and also phase 
mixing process of unbound recent groups) in a same cell of the phase space. Then, a {\em supercluster}-like 
velocity structure do not need any physical process to be maintained. At a given time 
in the solar neighbourhood, the juxtaposition at random of several cluster remnants creates over densities in 
the velocity field which mimics the existence of a physical entity with large velocity dispersion.
\subsection{Hipparcos sample}
Hipparcos data \cite{Hip97}, providing accurate distances and proper motions for complete
volume limited samples of nearby stars, offer the first opportunity to look at the phase mixing process and 
the disc heating mechanisms in action. Such signatures are searched for with a sample
that provides all the stars with same physical properties within a well defined volume. 
The sample was pre-selected inside the Hipparcos Input Catalogue (\cite{Hic92}) among the ``Survey stars''. 
The limiting magnitude is $m_{v} \leq 7.9 + 1.1\cdot\sin\mid b\mid$ for spectral types 
earlier than G5 (\cite{Tur86}). 
Spectral types from A0 to G0 with luminosity classes V and VI were kept. Within this 
pre-selection the final choice was based on Hipparcos magnitude ($m_{v} \leq 8.0$), 
colours ($-0.1\leq B-V \leq 0.6$), and parallaxes ($\pi \geq 8 mas$). The sample studied 
(see sample named h125 in Paper I, \cite{Creze98}) is a slice in absolute magnitude of this selection 
containing 2977 A-F type dwarf stars with absolute magnitudes brighter than 2.5. It is complete within 
125 pc from the Sun.
\subsection{Where to find...}
The wavelet analysis procedure is described and discussed in Section \ref{sec:implement}. Age determination  
methods follow in Section \ref{sec:agedet}. Main results of the density analysis in position space (clustering) are 
given in Section \ref{sec:density}. The analysis of the velocity space (streaming) is given in 
Section \ref{sec:velocity}. A critical review of uncompleteness and other systematic effects can be found in Paper III, 
Sections 5.1 and 5.2.
Section \ref{sec:howeggenscl} presents our understanding of the {\it supercluster} concept based on all the results 
detailed in Paper III. Conclusions in 
Section \ref{sec:conclu} present a simple scenario which organizes the main results of this paper to explain 
observed phase space structures in the solar neighbourhood.\\
%
\section{Density-velocity analysis using wavelet transform}
\label{sec:implement}
An objective method should first be adopted to identify structures and determine their characteristic scales and 
amplitudes. The wavelet transform does provide such an accurate local description of signal characteristics 
by projecting it onto a basis of compactly supported functions (\cite{Daub91}). The basis of wavelet functions 
is obtained by dilatation and translation of a unique, oscillating and zero integral function: the {\em mother wavelet}. 
The wavelet representation gives signal characteristics in terms of location  both in position 
and scale simultaneously.  
 \subsection{The wavelet transform}
The wavelet transform by $\Psi(x)$ of a real one-dimensional signal $F(x)$ is defined as the scalar product:\\
\begin{equation}
W_{s}(i)=\frac{1}{\sqrt{s}}\int^{+\infty}_{-\infty}F(x)\cdot \Psi^{\ast}(\frac{x-i}{s})dx\\
\end{equation} 
where $s$ is the scale and $i$ the position of the analysis.
Shape and properties of the so-called {\em mother wavelet} $\Psi(x)$ are similar to a Mexican Hat and ensure a 
quasi-isotropic wavelet transform of the signal. It is constructed from a $B_{3}$ spline function $\Phi(x)$ which is 
compact and regular up to the second order derivative:
\begin{eqnarray}
\Phi(x)=( \mid x-2\mid^{3} - 4\mid x-1\mid^{3} + 6\mid x\mid^{3} \nonumber\\
\hspace{1.cm} - 4\mid x+1\mid^{3}+\mid x+2\mid^{3} )/12
\end{eqnarray}
leading to:
\begin{equation}
\Psi(x)=\Phi(x)-\frac{1}{2}\Phi(\frac{x}{2})\\
\label{eq:relation}
\end{equation} 
The 3-D {\em  scaling function} $\Phi(x,y,z)$ used to analyze the observed $F(x,y,z)$ distribution is a separable function
\begin{equation}
\Phi(x,y,z)=\Phi(x)\cdot\Phi(y)\cdot\Phi(z)\\
\end{equation} 
Among several possible implementations of the wavelet analysis, the {\it ``\`a trou''} algorithm, 
previously used for the analysis of large scale distribution of galaxies (Lega et al, 1996, Bijaoui et al, 1996) 
has been selected. The reasons for this choice and the principles of the method are given in the following 
Section.\\
 \subsection{The ``\`a trou'' algorithm}
\label{sec:atrou}
Two main lines of algorithms have been used in order to implement the wavelet analysis concept. The pyramidal
approach (\cite{Mallat89}), used by Meneveau (1991) for turbulence analysis, with orthogonal wavelet 
basis addresses orthogonality problems in a rigorous way but it fails providing convenient tools to localize 
structures. On the contrary, the ``\`a trou'' algorithm associated with non orthogonal wavelet basis, 
giving an identical sampling at each scale (same number of wavelet coefficients at each scale), is appropriate 
for such practical purpose. For a detailed description of the ``\`a trou'' algorithm the reader is referred 
to  \cite{Holschneider89}, \cite{Starck93}, Bijaoui et al (1996) and details are also provided in Paper III.\\
The 3-D distribution $F(x,y,z)$ is binned in a 128 pixel edge cube which is wavelet analyzed 
on five dyadic scales. The analyzing function is dilated so that the distance between two bins increases by 
a factor 2 from scale $s-1$ to scale $s$. So it is a suitable framework in which local over-densities of 
unknown scales and low amplitude can be pointed at. The wavelet coefficient value depends on the 
gradient of the signal $F(x,y,z)$ in the neighbourhood of point $(i,j,k)$ considered at a given scale $s$: 
the absolute value of the coefficient increases all the more since the signal varies on this scale. 
For a positive or null signal like the observed distributions (star counts in the solar neighbourhood), 
over-density structures mainly lead to strong positive values of wavelet coefficients while under-densities 
cause strong negative coefficients.\\
\subsection{Thresholding and segmentation}
\begin{sloppypar}
Once the wavelet decomposition of the real signal is obtained, significant wavelet coefficient structures are 
separated from those generated by random fluctuations of a uniform background. A thresholding is applied at 
each scale in the space of wavelet coefficients. Thresholds are set at each scale and each position 
by estimating the noise level generated at the same scale by a uniform random signal built with the 
same gross-characteristics as the observed one at the position considered. Then a segmentation 
procedure returns pixel by pixel the characteristic extent of structures in each dimension. 
The thresholding and segmentation procedures are sketched out in Paper III.\\
\end{sloppypar}
Since we are aiming at an estimation of the total fraction of stars actually involved in physical 
structures, one has to care for the casual presence of background stars at the position (or 
velocity) of structures. Cluster membership is classically tested by the coincidence  of tangential 
velocities (Section \ref{sec:implementdens}). This is not possible for stream membership. 
In this last case, only the fraction of non-members can be evaluated (Paper III, Section 5.2). Eventually 
the  age distribution of members is discussed in the light of various possible scenarii.\\
%
%
\section{Individual age determination}
\label{sec:agedet}
\subsection{Introduction}
In order to bridge the observed phenomenology with the galaxy evolution, it is 
essential that structures in the phase space be dated. Once structure members have been duly 
identified  they should be given an age. Str\"omgren photometry is available  for some 1608 stars 
out of a total sample of 2977. When Str\"omgren photometry is available, ages are estimated in a 
now classical way (Section \ref{sec:agestrom}). There is however a strong suspicion that the age 
distribution  of stars observed in Str\"omgren photometry is biased. At least, even though observers 
are not likely to have selected their targets on a prior age indicator, they are likely to have favoured 
stars in clusters and streams, which is highly damageable for the present investigation. In order to 
correct at least statistically the suspected resulting bias, we propose an empirical palliative age 
estimation method based on the Hipparcos absolute magnitude and colour. This method is fully 
described in Paper III.
\subsection{Ages from Str\"omgren photometry}
\label{sec:agestrom}
\begin{figure}
  \begin{center}
  \centerline{\epsfig{file=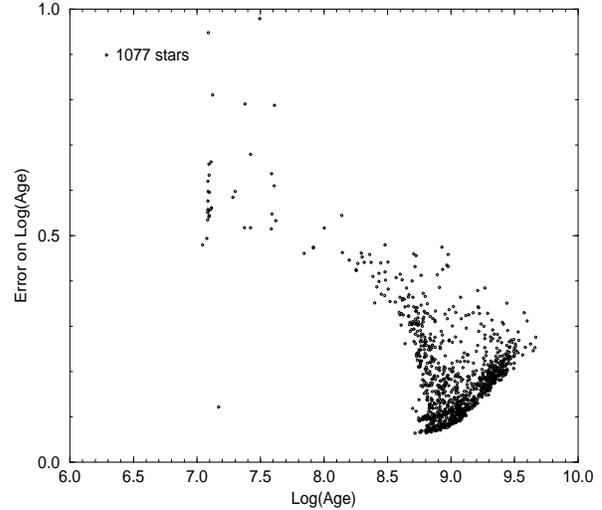,width=8.cm,height=9.cm,angle=-90.}}
  \end{center}
  \caption{\em Str\"omgren age determinations and their errors for 1077 stars. }
  \label{fig:agerror}
\end{figure}

\begin{sloppypar}
We find 1608 stars with published Str\"omgren photometry (Hauck \& Mermilliod, 1990) in our sample. 
For this sub-sample, the effective 
temperature $T_{eff}$, the surface gravity $\log g$ and their errors, the metallicity [Fe/H] are derived 
from the photometric indices $(b-y)$, $m1$, $c1$, $H_{\beta}$, the visual magnitude $m_{v}$ and the rotational 
velocities $v\cdot\sin i$ (if published in \cite{Uesugi81}) using a program developed by E. 
Masana (\cite{Figue91}, \cite{Mas94}, 
\cite{Jordi97}). Based on the above three physical parameters and a model of stellar evolution taking into account 
overshooting effects and mass loss (Schaller et al, 1992), ages and masses are inferred 
with the code developed by Asiain (Asiain et al, 1997). By means of the stellar metallicity, this algorithm 
interpolates the set of stellar evolutionary models to work at the appropriate metallicity. Ages and masses 
determination were possible for only 1077 stars (a third of the sample) due to the failure of the method to 
get reliable metallicities for spectral types between A0 and A3.\\
\end{sloppypar}
The mean error on age determination is 30 $\%$ ($\sim$0.2 in logarithmic scale) for the bulk of the stars 
(Figure \ref{fig:agerror}) and rarely exceed 60 $\%$ ($\sim$0.3 in logarithmic scale). Only the youngest stars, 
between 10$^{7}$ and 10$^{8}$ years, have errors above 100 $\%$ ($\sim$0.5 in logarithmic scale). Nevertheless, this 
precision is sufficient to unambiguously attach them to the youngest age group. 
The distribution of  Str\"omgren ages (Figure \ref{fig:agedist1}) for these 1077 stars ranges 
from 10 Myr to 3 Gyr with a peak around 650 Myr ($\log$(age) = 8.8).
 \subsection{Palliative ages from $(M_{v},(B-V))$}
 \label{sec:agemvbv}
On a first step, we use existing ages to draw a plot of ages versus (Mv, (B-V)). A primary age parameter 
is assigned as the mean age associated to a given range of (Mv, (B-V)). 
Then Str\"omgren age data are used a second time to assign a probability distribution of  palliative ages 
as a function of the primary age parameter (See details in Paper III).
  \begin{figure}
   \vspace{-1cm}
   \begin{center}
   \epsfig{file=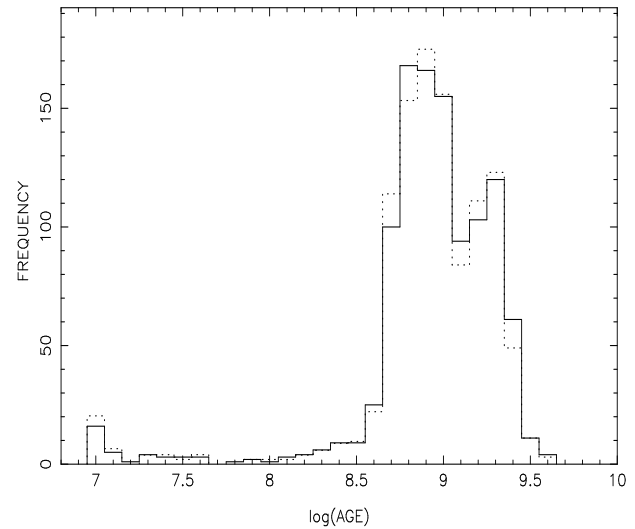,width=8.cm,height=9.cm,angle=-90.}
   \end{center}
   \caption{\em Str\"omgren age distribution obtained with 1077 stars (full line) and palliative age distribution for the same sub-sample (dotted line).}
   \label{fig:agedist1}
 \end{figure}
  \begin{figure}
   \vspace{-1cm}
   \begin{center}
   \epsfig{file=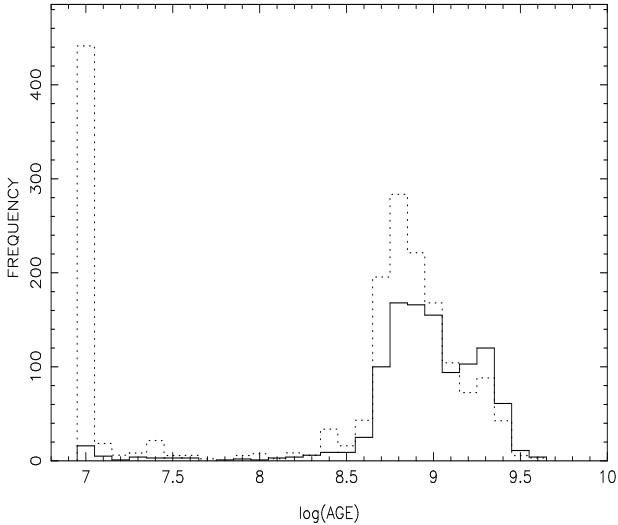,width=8.cm,height=9.cm,angle=-90.}
   \end{center}
   \caption{\em Str\"omgren age distribution obtained with 1077 stars (full line) and palliative age distribution for the remainder, i.e stars without Str\"omgren photometry (dotted line).}
   \label{fig:agedist2}
 \end{figure}
This process produces a palliative age distribution which is free from possible biases affecting the 
sub-sample of stars with Str\"omgren photometry for stars with B-V $\geq$ 0.08. \\
The palliative age distribution of stars with Str\"omgren photometry is presented on 
Figure \ref{fig:agedist1}. The small deviations from the distribution of original Str\"omgren 
ages are due to finite bin steps used in the discretisation process. The palliative age 
distribution of the sub-sample without Str\"omgren photometry  (Figure \ref{fig:agedist2}) 
shows a great difference for very young ages with respect to the Str\"omgren 
age distribution. The great peak at $10^{7}$yr means that more than 400 stars are clearly younger than 
2-3 $10^{8}$ yr (see explanations in Paper III) and compose a separate population from the rest of the sample. \\
These palliative ages have only statistical significance. Hence, they sometimes produce artifacts or young 
ghost peaks in some stream age distributions. Such dummy young peaks will always appear as the 
weak counterpart of a heavy peak around $log(age)=8.7$. Nevertheless palliative ages permit 
to shed light on the age content of the phase space structures when Str\"omgren data are sparse.\\
%
%
\section{Clustering}
\label{sec:density} 
\subsection{Searching for clusters}
\label{sec:implementdens}
\begin{figure*}
  \begin{center}
   \epsfig{file=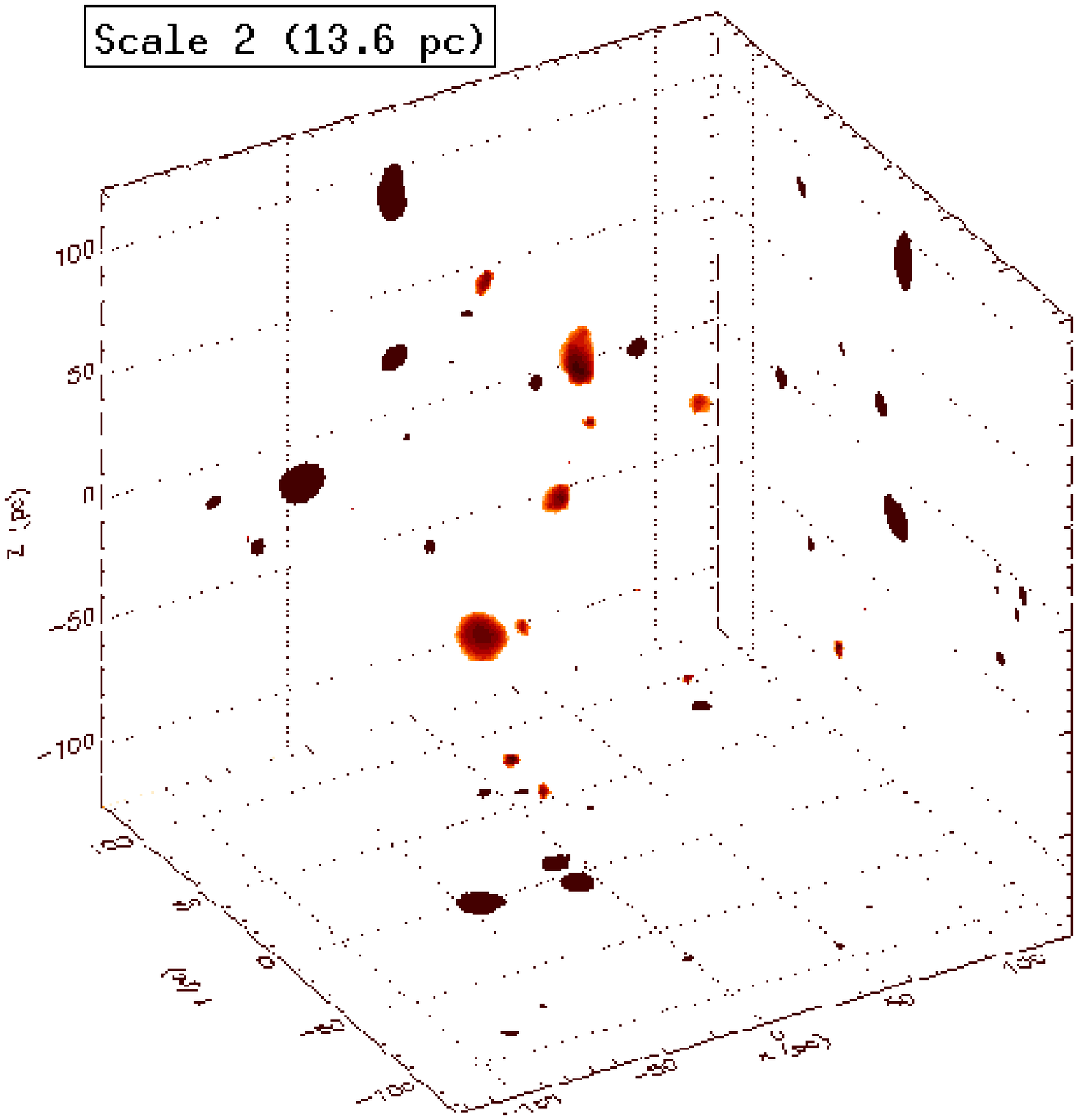,height=8.cm,width=8.cm,angle=0.} 
  \epsfig{file=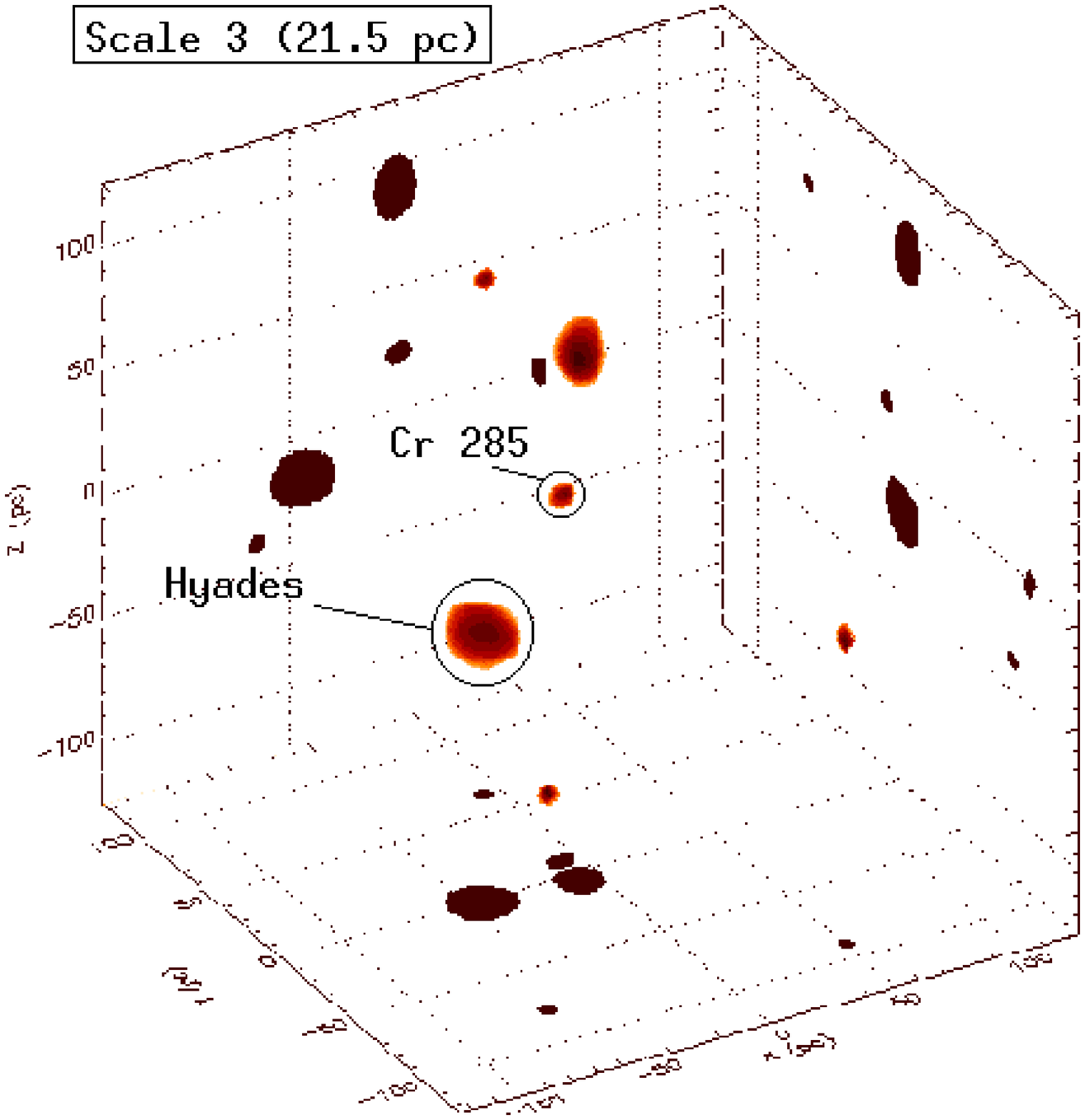,height=8.cm,width=8.cm,angle=0.}
  \epsfig{file=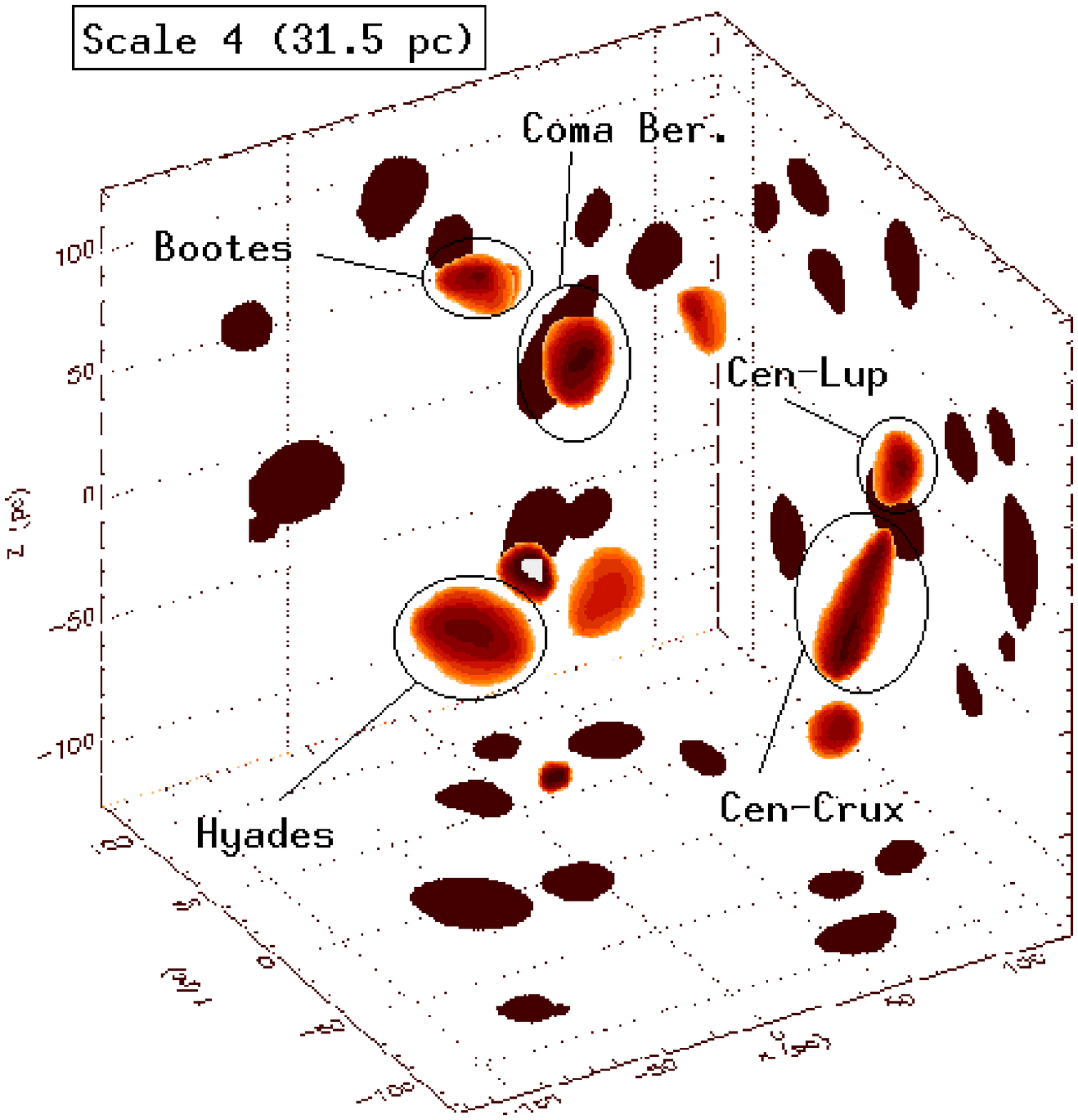,height=8.cm,width=8.cm,angle=0.}
  \epsfig{file=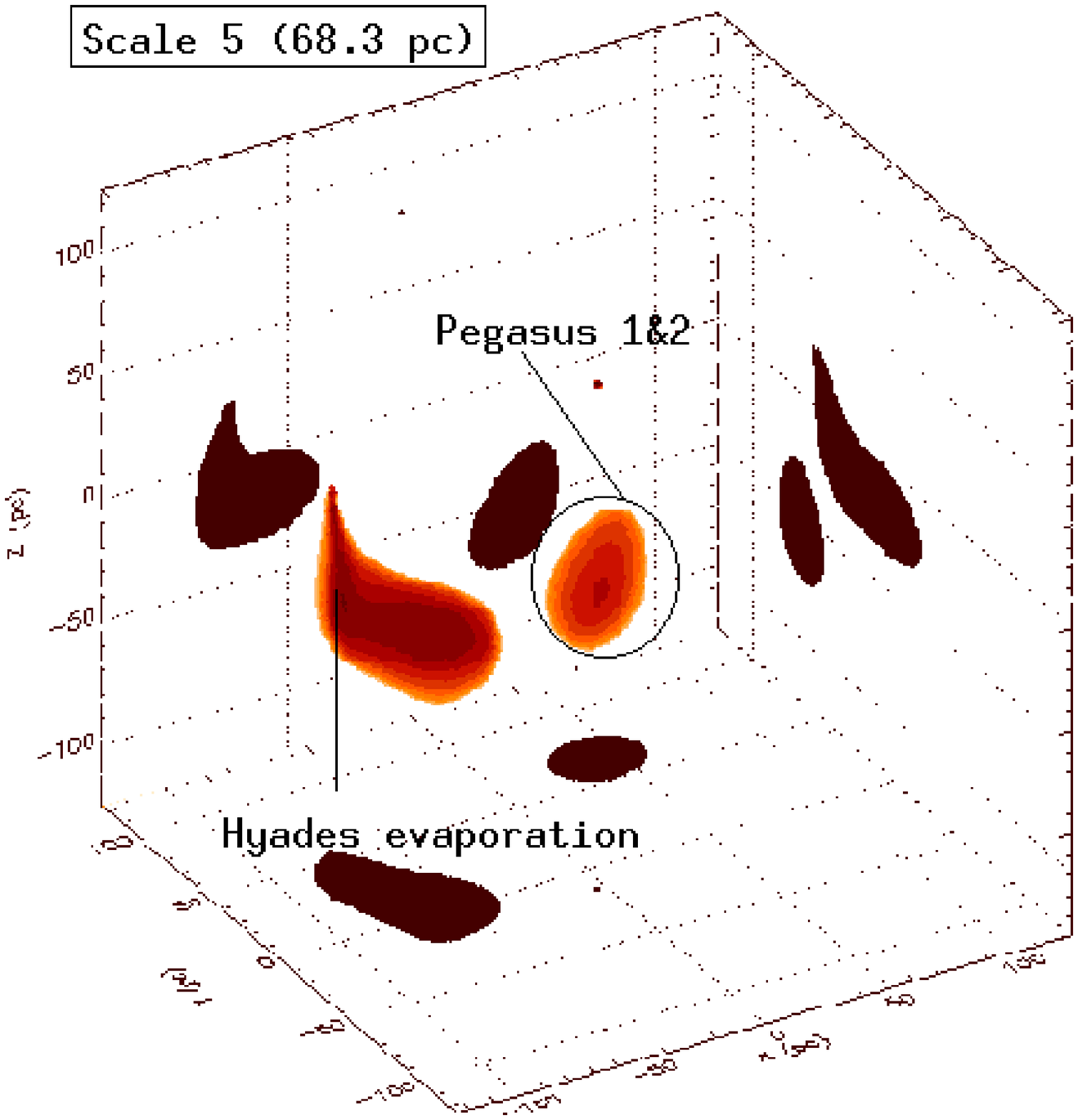,height=8.cm,width=8.cm,angle=0.}
  \caption{\em Significant over-densities in wavelet coefficient spaces for space distributions (grey clumps) and their projections on the three main planes (black clumps): (X,Y) plane (bottom) - (Y,Z) plane (right) - (X,Z) plane (background). Some cluster identifications are given. Scale 2, 3, 4 and 5.}
  \label{fig:xyzwave}
  \end{center}
\end{figure*}

(X,Y,Z) distributions range from -125 pc to +125 pc and are binned in a Sun centered orthonormal 
frame, X-axis towards the galactic center, Y-axis in the rotation direction and Z-axis towards the 
north galactic pole. The discrete wavelet analysis is performed on five scales: 9.7, 13.6, 21.5, 37.1 
and 68.3 pc. These values correspond to the size of the dilated filter at each scale.\\
Over-densities are identified at each scale (Figure \ref{fig:xyzwave}) by the 
segmentation procedure and the stars belonging to each volume are collected. Due to the over-sampling 
of the signal by the ``\`a trou'' algorithm, some structures are detected  on several scales. A cross-correlation 
has been done between all scales to keep the structure at the largest scale provided that there is no 
sub-structure at a lower scale (higher resolution).  An iterative 2.5 sigma clipping procedure on tangential 
velocity distributions for each group remove field stars and select structures with coherent kinematics.\\
The same work has been performed to search for under-densities (signed by negative 
wavelet coefficients). No convincing evidence of the presence of void has been found. In the following 
we only pay attention to the detection of over-densities.\\
\subsection{Main results}
\label{sec:clustdet}
\begin{figure}
  \begin{center}
  \epsfig{file=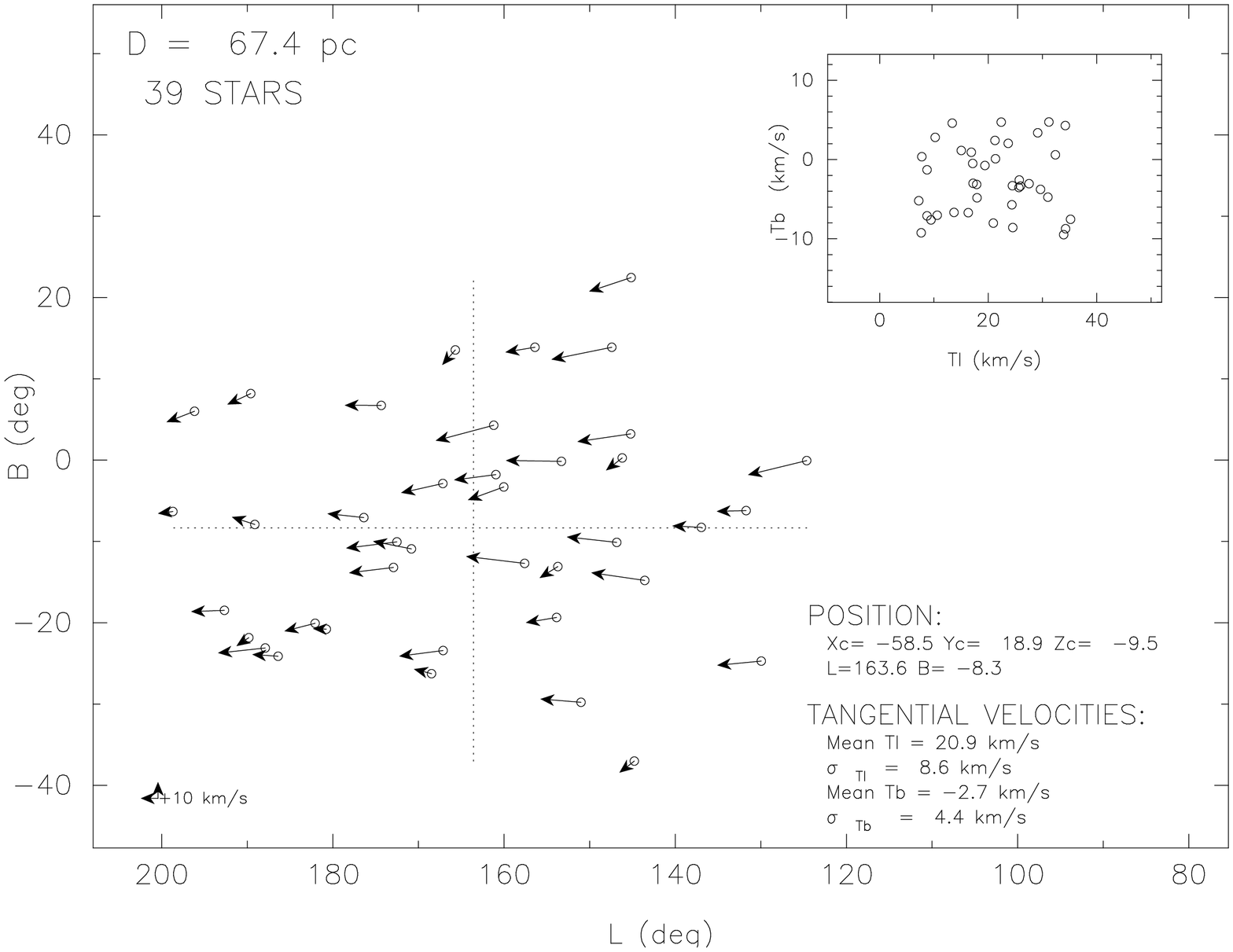,width=7.cm,height=8.cm,angle=-90.}
  \epsfig{file=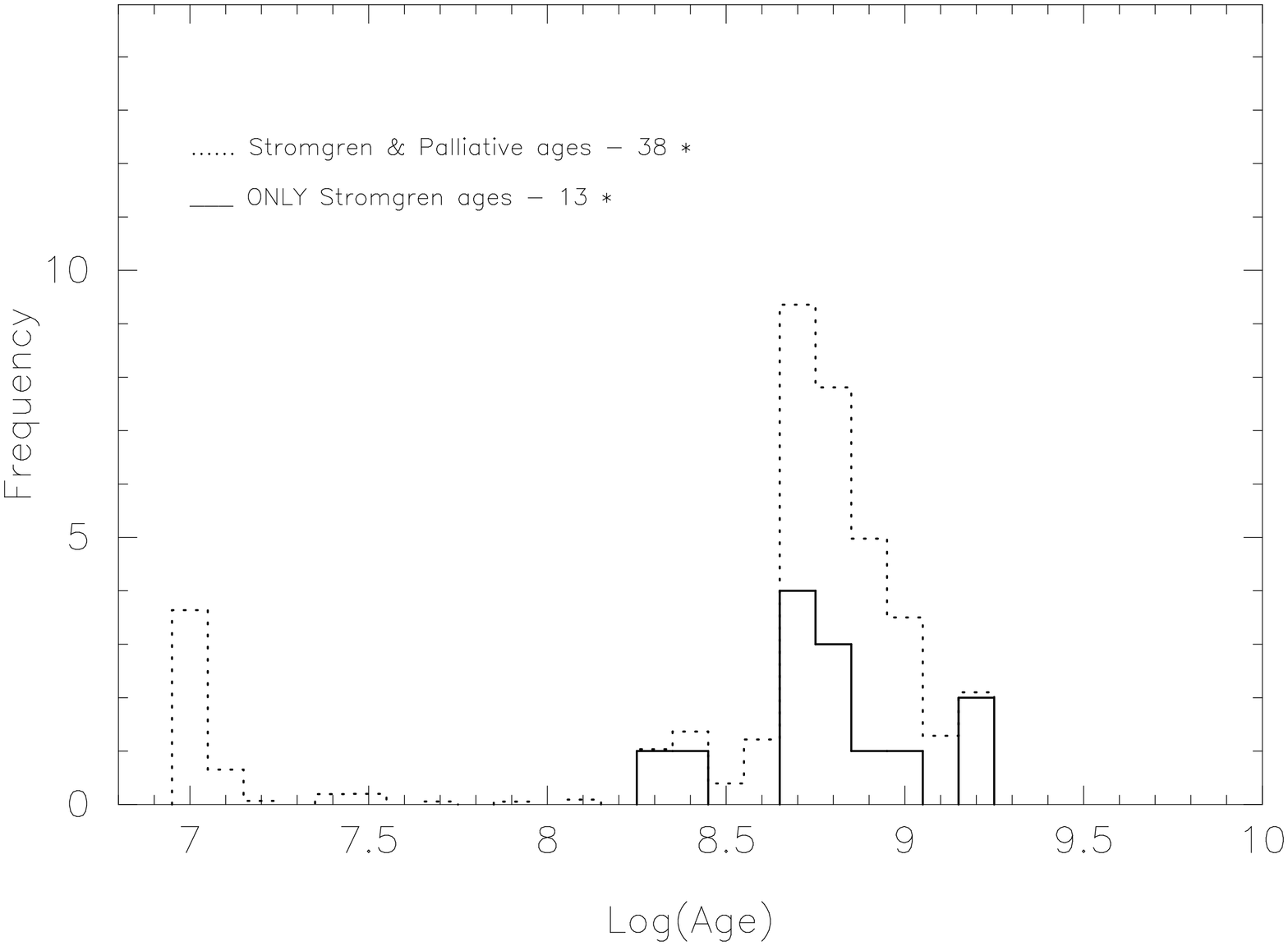,width=7.cm,height=8.cm,angle=-90.}
  \epsfig{file=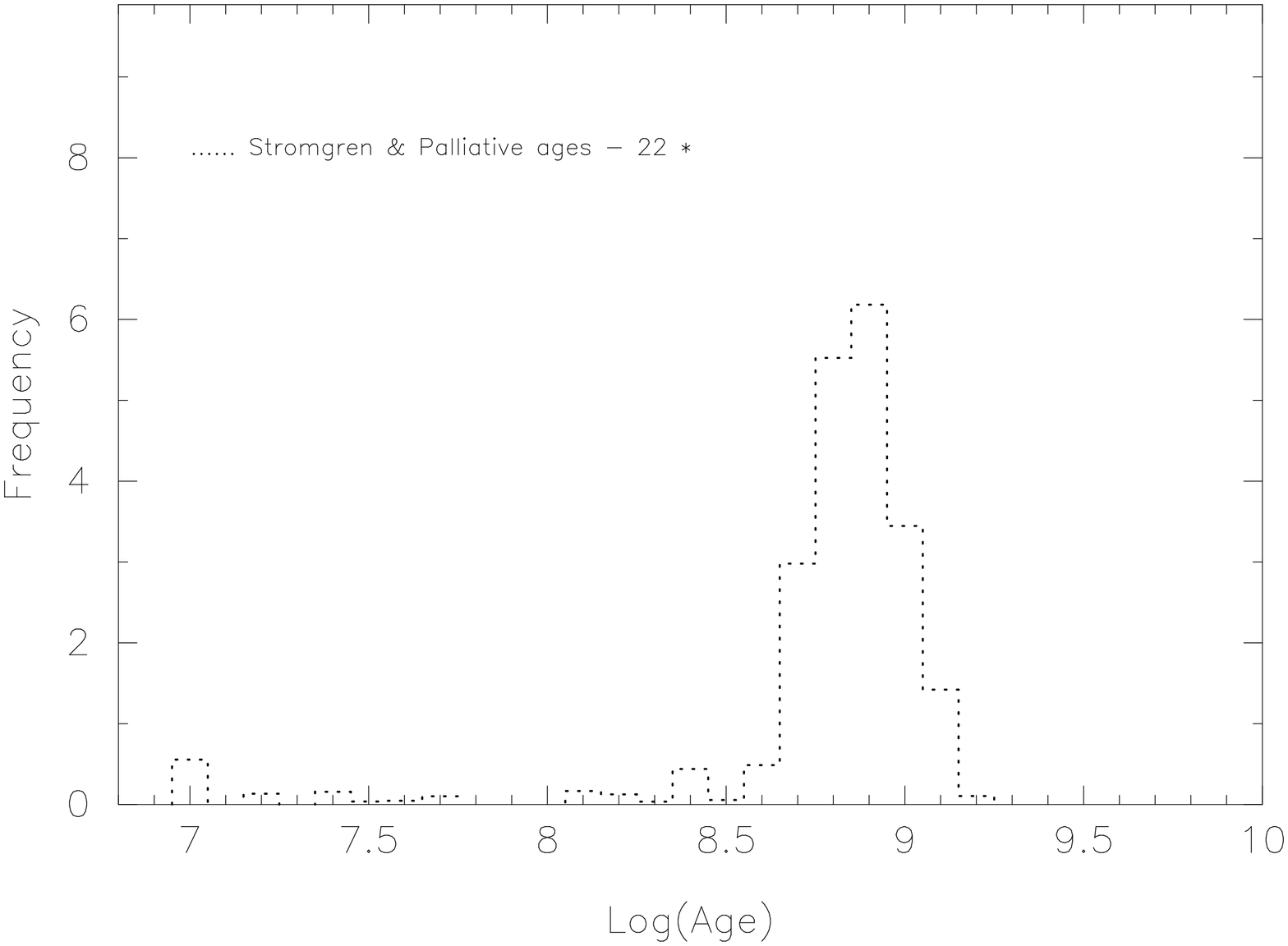,width=7.cm,height=8.cm,angle=-90.}
  \end{center}
  \vspace{-0.3cm}
  \caption{\em The Hyades open cluster's tail after selection on tangential velocities ({\bf top}) and its age distribution ({\bf middle}), Hyades OCl members age distribution ({\bf bottom}).}
  \label{fig:tail}
 \end{figure}

The space distribution is essentially smooth at all scales. The volumes selected by the segmentation 
procedure contain 10 per cent of the stars. After the 2.5 sigma clipping procedure on the tangential 
velocities, only 7 per cent are still in clusters or groups. Most of them are well known: Hyades, 
Coma Berenices, Ursa Major open clusters (hereafter OCl) and the Scorpio-Centaurus association. 
Otherwise, three new groups, probably loose clusters, are detected: Bootes and Pegasus 1 and 2.
Paper III provides a detailed review of all these cluster characteristics. Here, we just focus on a newly 
discovered feature concerning the Hyades OCl: its evaporation track.\\
\\
{\bf The Hyades open cluster's tail} is clearly visible at scale 5 on Figure \ref{fig:xyzwave}. 
After selection on tangential velocities, 39 stars of the over-density tail are found to have similar 
motions (Figure \ref{fig:tail}). The tangential velocity component along the $l$ axis is nearly the 
same as the Hyades OCl's one (20.8 $km\cdot s^{-1}$ vs. 19.4 $km\cdot s^{-1}$ respectively) but 
differs along the $b$ axis (-2.6 $km\cdot s^{-1}$ vs. 14.2 $km\cdot s^{-1}$ respectively). 
The extremely peaked age distribution at $5-6.3\cdot 10^{8}$ yr shows that stars are slightly 
younger on average than the Hyades' ones ($6.3-8\cdot 10^{8}$ yr). Both age determinations are 
in agreement with recent age determination by \cite{Perry98} who give $6.25\pm 0.5 \cdot 10^{8}$ yr.
Nevertheless, the density distribution along the tail are a spectacular confirmation of the 
theoretical predictions proposed by Weidemann et al (1992) concerning the distributions of 
evaporated stars from the Hyades OCl. These authors expect stars evaporated from the Hyades 
to be distributed in a {\it needlelike ellipsoid centred on the cluster center and with longest axis pointing 
towards $(l=97.7^{o}, b=+1.7^{o})$}. The space distribution of the 39 stars exhibit an obvious major axis, 
with respect to the Hyades OCl, pointing towards $(l=94.0^{o}, b=+7.0^{o})$ but are only distributed 
further forward in the direction of Galactic rotation. This feature can be produced by escaping stars 
orbiting closer to the Galactic center than the cluster members (Weidemann et al, 1992). These stars have 
a smaller guiding radius implying a shorter rotation period. The authors speculate for this type of escaping 
stars, the existence of a phase advance in the vertical oscillation which could agree with the difference 
observed in the $T_{b}$ components. Moreover, we observe these stars higher in the plane as it is expected 
because the open cluster is on the upswing.\\
It is striking to notice that stars as massive as 1.8$M_{\odot}$ (which is the typical mass in our sample) 
are also evaporating from the cluster. Whatever the cause of the ``evaporation'' process, slow random change 
of star binding energies due to a large number of weak encounters (Chandrasekhar, 1942) or sudden energy 
increase by few close encounters (H\'enon, 1960) inside the cluster, it produces a mass segregation among still 
clustered stars. Indeed, low mass stars are preferentially evaporated and massive stars are retained as 
shown by means of numerical simulations in \cite{Aar73} and \cite{Terlev87} and as reported in 
\cite{Reid92} who find a steeper density gradient among the brighter Hyades stars. The asymmetry 
in the density distribution of the escaping stars could have two main explanations. We cannot rule out a 
non detection by the wavelet analysis of a symmetric tail provided that its size is larger than the coarser scale 
of analysis. But if it were not the case, it might sign an encounter between the Hyades OCl and a 
massive object on the Galactic center side of the cluster. This hypothesis was previously investigated by 
\cite{Perry98} who find it highly improbable because of the velocity with respect to the LSR 
(30 $km\cdot s^{-1}$) and the mass ($\sim 10^{6} M_{\odot}$) needed for such an object. 
A more detailed analysis should be done to conclude on the origin of this tail.\\
%
%
\section{Streaming} 
\label{sec:velocity} 
\subsection{Mean velocity field}
\label{sec:ellipsoid}
Global characteristics of the triaxial velocity ellipsoid are obtained from a sub-sample of 1362 stars 
which have observed and published radial velocities (\cite{Hic92}). The centroid and the velocity dispersions in the 
orthonormal frame centred on the Sun's velocity with U-axis towards the galactic center, V-axis 
towards the rotation direction and W-axis towards the north galactic pole, are the following:\\
\begin{itemize}
\item $\overline{U}_{sample}$ = -10.83 $km\cdot s^{-1}$  and $\sigma_{U}$= 20.26 $km\cdot s^{-1}$
\item $\overline{V}_{sample}$ = -11.17 $km\cdot s^{-1}$ and $\sigma_{V}$= 12.60 $km\cdot s^{-1}$
\item $\overline{W}_{sample}$ =  -6.94 $km\cdot s^{-1}$ and $\sigma_{W}$= 8.67 $km\cdot s^{-1}$\\
\end{itemize}
The sub-sample with observed radial velocities is incomplete and contains biases since the stars are not observed at 
random. To keep the benefits of the sample completeness, a statistical convergent point method is developed to 
analyze all the stars (see Paper III, Section 5.1).
\subsection{Wavelet analysis of the velocity field}
In a recent paper \cite{Deh98} criticizes this methodology which could be less rigorous than 
maximizing the log-likelihood of a velocity distribution model. 
The author argues that direct determination of velocity dispersions 
based on a convergent point method produces overestimation and may create
spurious structures on small scales by noise amplification.
Both risks are clearly ruled out by the calibration process described above :\\
\begin{enumerate}
\item the thresholding is calibrated on numerical experiments so as 
to exclude spurious structure detection at any meaningful level (see Paper III, Section 2.2).
\item the fraction of spurious members in real groups created by the
convergent point method is estimated from the available subset of true
3D velocities based on observed radial velocity data (see Paper III, Section 5.2.1). 
\item estimated group velocity dispersions are derived from radial 
velocity subset as well, excluding convergent point reconstructed data. 
\end{enumerate}
The confusion in Dehnen's comment is probably related to another
misunderstanding. In a footnote dedicated to a preliminary version
of our work, this author explicitly suggests that the convergent point
method ``{\em which stems for the times when more rigorous treatment was
impractical on technical ground}'' was adopted by plain anachronism or
ignorance of ``{\em more rigorous treatment}''. Actually, the choice was guided
by the requirements of the wavelet analysis which gives access to a fine 
perception of 3D structures. Dehnen's analysis fits {\em more rigorously} a velocity 
field model with an initial coarse resolution of 2-3 $km\cdot s^{-1}$ depending on the 
velocity component considered while our bin step is 0.8 $km\cdot s^{-1}$.
As a result we get (calibrated) significant signal one scale below the finest resolution 
of Dehnen's work. Moreover, the significance threshold of velocity field features obtained 
by Dehnen is somewhat arbitrary: features are relevant if they appear in more than one of the 
studied sub-samples.\\
Velocities of open cluster stars are poorly reconstructed by this method because their members are spatially close. 
For such stars, even a small internal velocity dispersion results a poor determination of the convergent point.  
For this reason, we have removed stars belonging to the 6 main identified space concentrations (Hyades OCl, 
Coma Berenices OCl, Ursa Major OCl and Bootes 1, Pegasus 1, Pegasus 2 groups) found in the previous spatial 
analysis. Eventually, the reconstruction of the velocity field is performed with 2910 stars.\\
\begin{figure}
 \begin{center}
    \epsfig{file=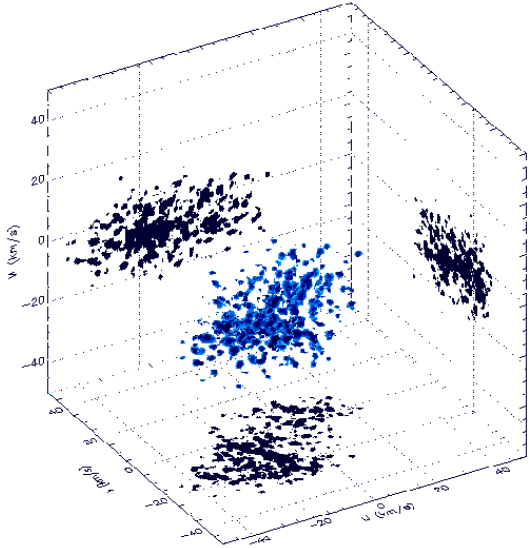,height=8.cm,width=8.cm,angle=0.} 
  \caption{\em Isosurfaces (grey clumps) in wavelet space after thresholding for velocity distributions - Projections (black clumps) on (U,V) plane (bottom), (V,W) plane (right) and (U,W) plane (background) - Scale 1 (3.2 $km\cdot s^{-1}$).}
  \label{fig:uvw_s1}
  \end{center}
\end{figure}
\begin{figure}
 \begin{center}
    \epsfig{file=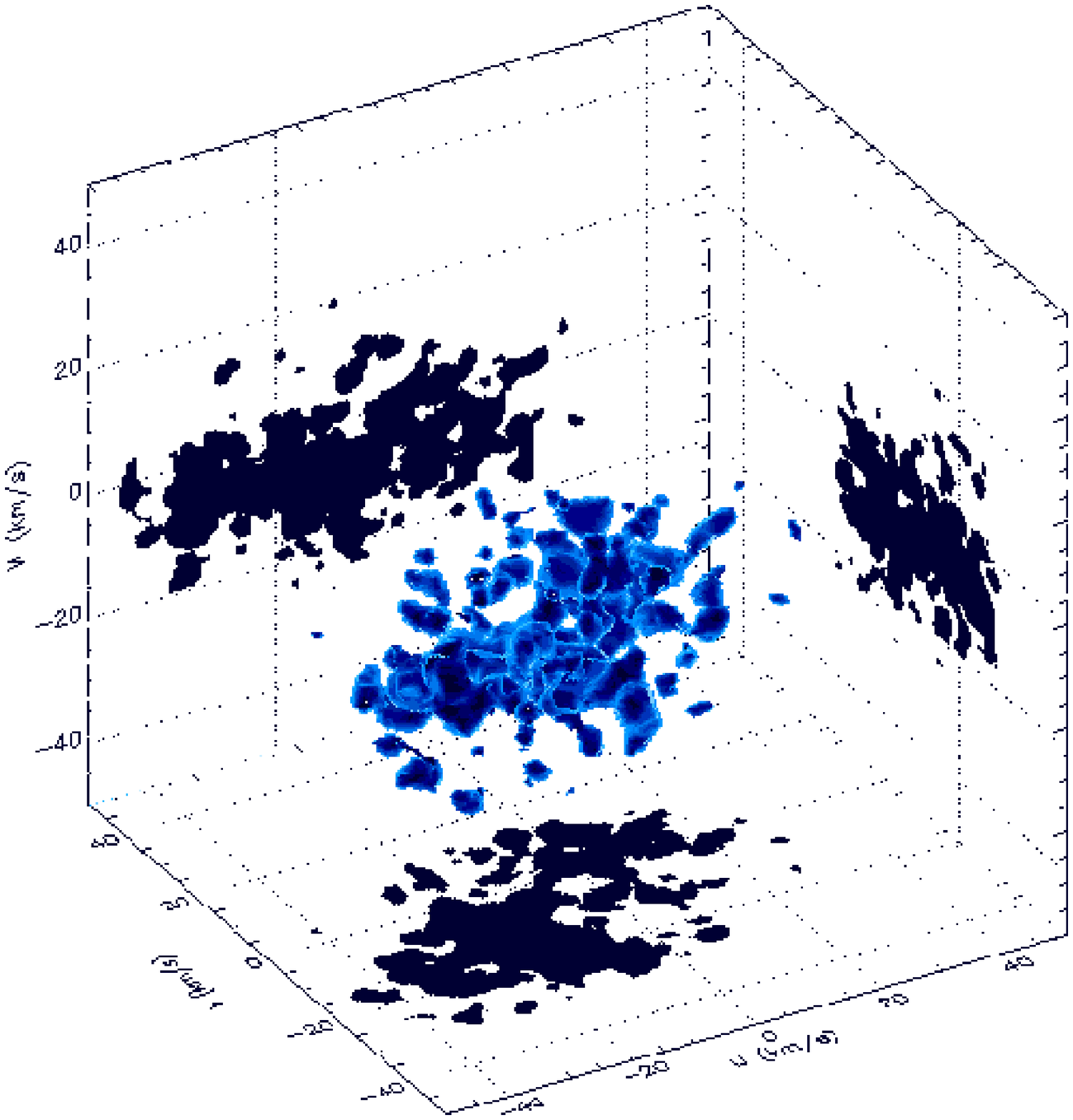,height=8.cm,width=8.cm,angle=0.} 
  \caption{\em Same as Figure \ref{fig:uvw_s1} for scale 2 (5.5 $km\cdot s^{-1}$).}
  \label{fig:uvw_s2}
  \end{center}
\end{figure}
\begin{figure}
 \begin{center}
    \epsfig{file=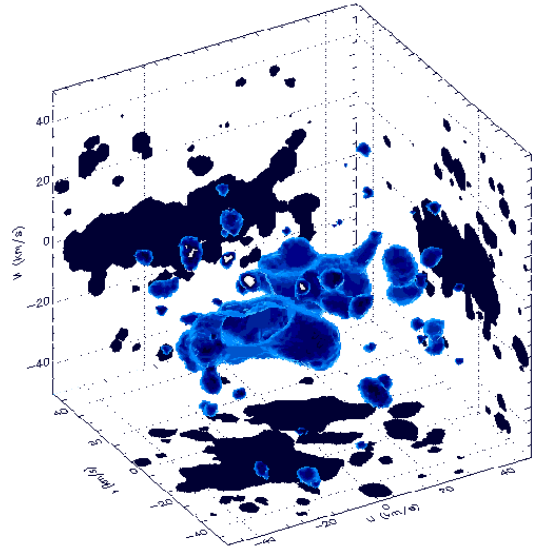,height=8.cm,width=8.cm,angle=0.} 
  \caption{\em  Same as Figure \ref{fig:uvw_s1} for scale 3 (8.6 $km\cdot s^{-1}$).}
  \label{fig:uvw_s3}
  \end{center}
\end{figure}
Reconstructed (U,V,W) distributions are given in an orthonormal frame centred in the Sun velocity 
(see Section \ref{sec:ellipsoid}) in the range [-50,50] $km\cdot s^{-1}$ on each component. The wavelet analysis 
is performed on five scales: 3.2, 5.5, 8.6, 14.9 and 27.3 $km\cdot s^{-1}$. In the following, the analysis focuses 
on the first three scales revealing the stream-like structures (see Figures \ref{fig:uvw_s1},  \ref{fig:uvw_s2} and 
\ref{fig:uvw_s3}), larger ones reach the typical size of the velocity ellipsoid. Once the segmentation procedure 
is achieved, stars belonging to velocity clumps are identify. Structures found in this reconstructed velocity field 
are contaminated by spurious members created by the method. This contamination is evaluated from the 
sub-sample with observed $V_{R}$ by a procedure described in Paper III, Section 5.2.1 (hereafter procedure 5.2.1). 
The proportion of field stars is also evaluated in Paper III, Section 5.2.2 (hereafter procedure 5.2.2).\\
\subsection{Stream phenomenology}
\label{sec:phenom}
\begin{figure}
 \begin{center}
    \epsfig{file=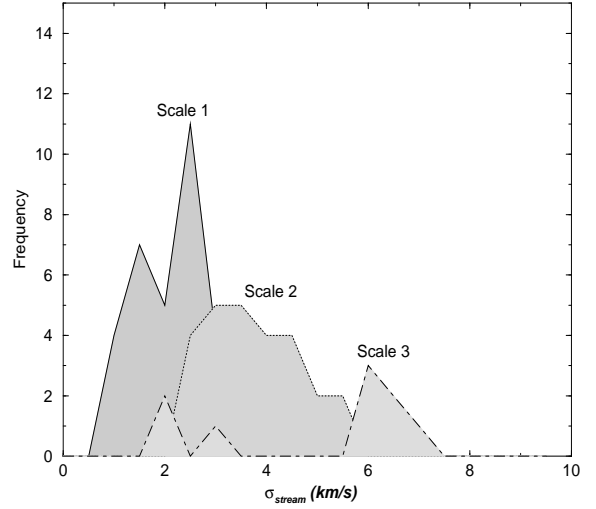,height=9.cm,width=8.cm,angle=-90.} 
  \caption{\em Distribution of stream mean velocity dispersions per scale. }
  \label{fig:distsigma}
  \end{center}
\end{figure}

The A-F type star 3D velocity field turns out highly structured at the first three scales. In a previous analysis 
\cite{Chereul97}, it has been found that structures are mainly revealed in the (U,V) plane rather than in the (U,W) plane. 
This is, probably, the signature of a faster phase mixing process along the vertical axis.\\
In Paper III, Tables 2, 3 and 4 provide mean velocities, velocity dispersions and numbers of stars remaining 
after correction procedures 5.2.1 and 5.2.2 for streams at respectively scale 3, 2 and 1. 
Here, Table \ref{tab:table2} summarizes characteristics of the streaming organization. For each scale, 
Table \ref{tab:table2} gives the number of detected velocity groups after wavelet analysis, the number of 
confirmed streams remaining after elimination of spurious members (procedure 5.2.1) and the fraction of 
confirmed stars in streams after corrections 5.2.1 and 5.2.2. The number of 
confirmed streams gives a lower estimation of the real number of streams since sometimes the selection 
of real members is performed on very few radial velocity data with respect to the potential members. 
We can notice that the fraction of stars involved in streams at scale 2 ($\sim$ 38 $\%$) is 
roughly the same as the fraction involved in larger structures at scale 3 ($\sim$46$\%$) when 
field stars are removed. It already gives a strong indication that large structures at scale 3 could be 
mainly composed by clustering of streams with smaller velocity dispersions from scale 2.\\
The wavelet scales can be calibrated a posteriori in terms of stream velocity dispersions. The distribution 
of mean velocity dispersions 
($\overline{\sigma}_{stream}=(\frac{1}{3}\cdot(\sigma_{u}^{2}+\sigma_{v}^{2}+\sigma_{w}^{2}))^{\frac{1}{2}}$) 
for the streams at a given scale yields a better indicator of the characteristic scale than the filter 
size (Figure \ref{fig:distsigma}). The typical velocity dispersions of detected streams at 
scales 1, 2 and 3 are respectively 2.4, 3.8 and 6.3 $km\cdot s^{-1}$.\\
Streams appearing at scale 3 ($\overline{\sigma}_{stream} \sim$ 6.3 $km\cdot s^{-1}$) correspond 
to the so-called Eggen {\em superclusters}. At smaller scales ($\overline{\sigma}_{stream} \sim$ 3.8 
and 2.4 $km\cdot s^{-1}$) {\em superclusters} split into distinct streams of smaller velocity dispersions.\\
\begin{table}
 \vspace{-0.1cm}
\caption{\em Number of streams detected per scale by the wavelet analysis and the percentage of stars involved in streams after cleaning up spurious members and field stars.}
  \label{tab:table2}
    \leavevmode
   \begin{center}
    \begin{tabular}[h]{lccc}
      \hline \\[-5pt]
	Scale &1 &2 &3 \\
      \hline \\[-5pt]
Filter size ($km\cdot s^{-1}$)&3.2&5.5&8.6\\
	&&&\\
$\overline{\sigma}_{stream}$ ($km\cdot s^{-1}$)& 2.4 &3.8 &6.3\\
	&&&\\
Detected velocity groups & & & \\
after wavelet analysis & 63& 46& 24\\
	&&&\\
Confirmed streams after elimination& & &\\
of spurious members (Procedure 5.2.1)& 38& 26& 9\\
	&&&\\
$\%$ of stream stars after elimination & & & \\
of spurious members (Procedure 5.2.1)& 17.9$\%$& 38.3$\%$& 63.0$\%$\\
	&&&\\
$\%$ of stream stars after elimination&&&\\
of background (Procedure 5.2.1 and 5.2.2)&17.9$\%$&38.3$\%$&46.4$\%$\\ 
    \hline \\[-5pt]
   \end{tabular}
   \end{center}
   \end{table}
The age distribution inside each stream is analyzed. The analysis is performed on three different data sets: 
\begin{itemize}
\item the whole sample (ages are either Str\"omgren or palliative),
\item the sample restricted to stars with Str\"omgren photometry data (without selection on radial velocity),
\item the sample restricted to stars with {\em observed } (as opposed to {\em  reconstructed}) radial velocity data 
(ages are either  Str\"omgren or palliative).
\end{itemize}
The selection on photometric ages gives a more accurate description of the stream age content while the 
last sample permits to obtain a reliable kinematic description since stream members are selected through 
procedure 5.2.1. All mean velocities and velocity dispersions of the streams are calculated 
with the radial velocity data set. Combining results from these selected data sets generally brings 
unambiguous conclusions. \\
Results obtained for all the streams found on these 3 scales are fully repertoried in Paper III. Here, we 
focus on the results for the Pleiades {\em supercluster} (hereafter Pleiades SCl). This example illustrates very well the advance realized in the 
understanding of {\em supercluster} inner structure.
\subsection{Inner structure of the {\bf Pleiades} {\em supercluster}}
\label{sec:pleiade_scl}
\begin{figure}
  \begin{center}
   \epsfig{file=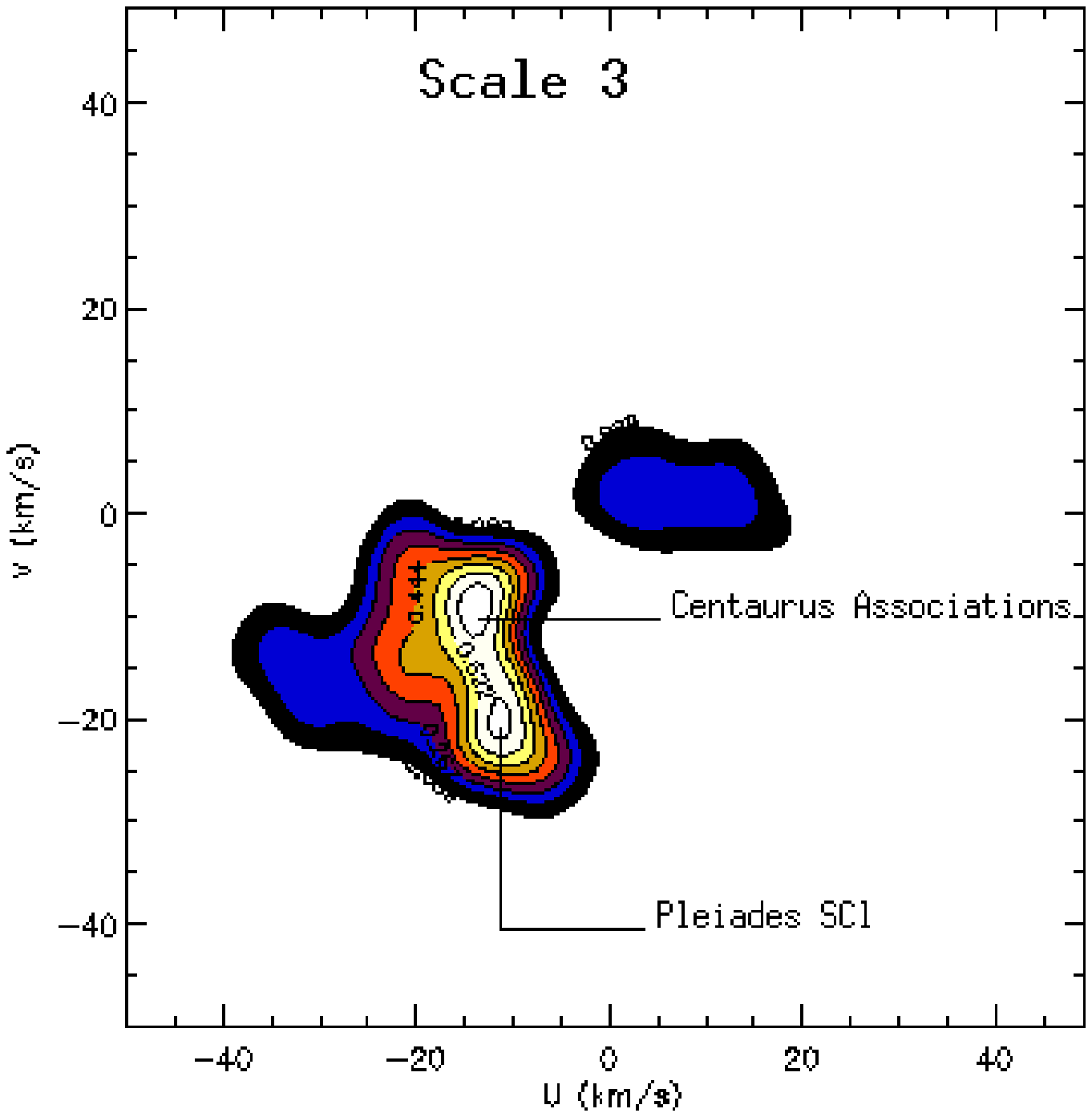,height=7.cm,width=7.cm,angle=0.} 
  \epsfig{file=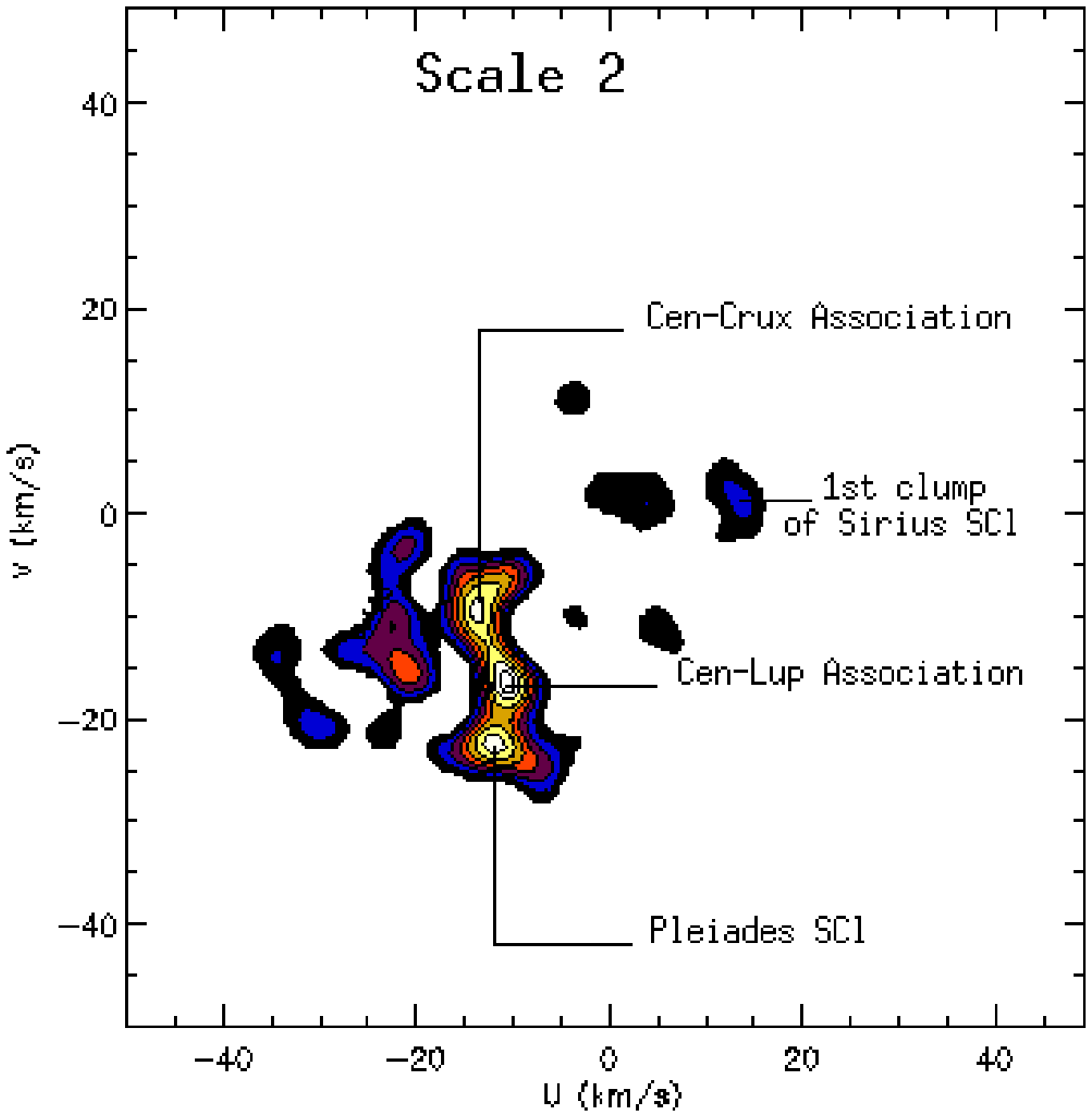,height=7.cm,width=7.cm,angle=0.}
  \epsfig{file=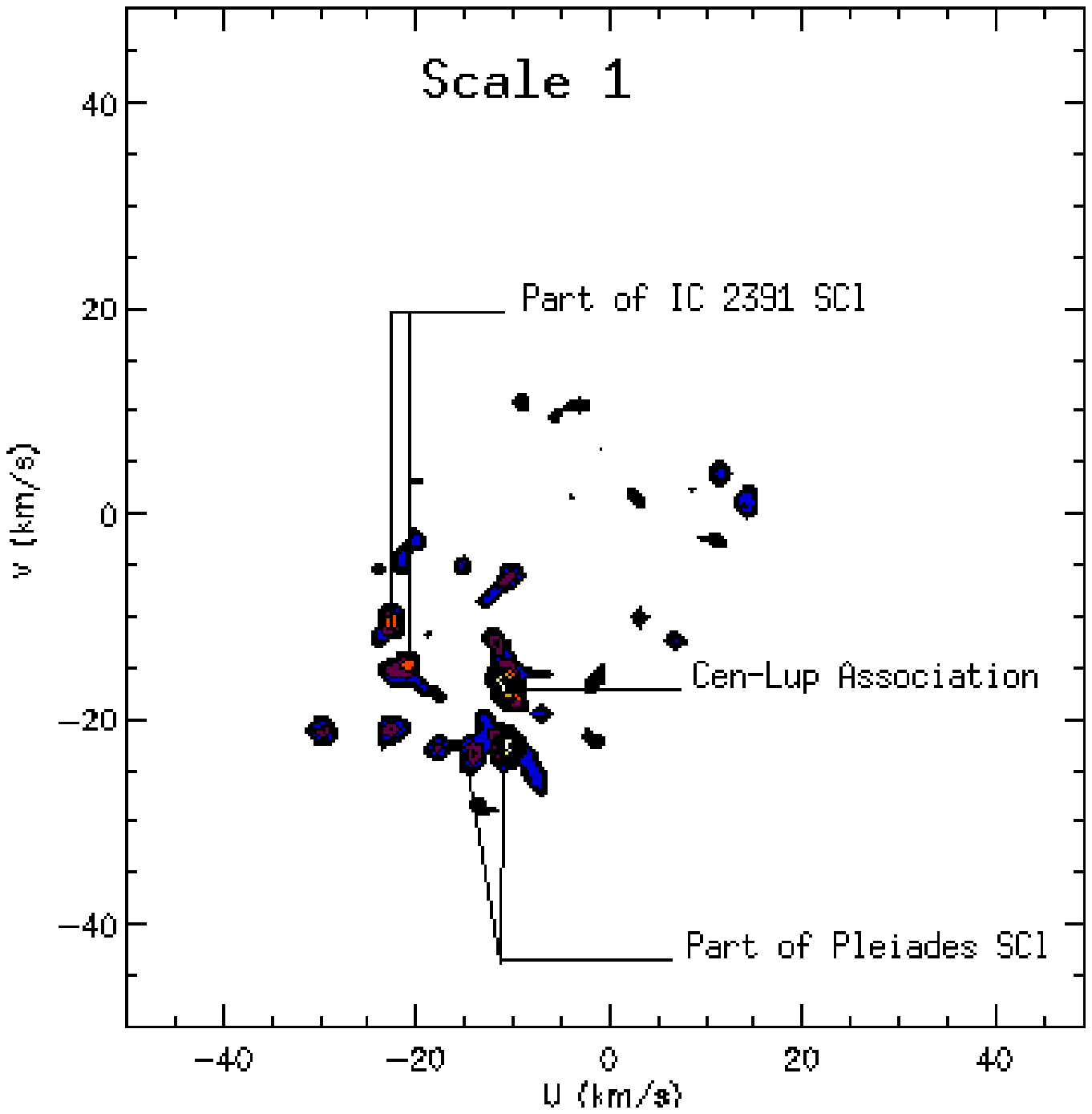,height=7.cm,width=7.cm,angle=0.}
  \caption{\em {\bf Pleiades SCl}. Thresholded wavelet coefficient isocontours at W=-6.2 $km\cdot s^{-1}$ of the velocity field at scale 3 ({\bf top}), scale 2 ({\bf middle}) and scale 1 ({\bf bottom}). Scale 2 shows the dichotomy between Cen-Crux and Cen-Lup associations (streams 2-26 and 2-12 in Paper III, Table 3) embedded in the same clump at scale 3. Scale 1 exhibits clearly the sub-structuring of the Pleiades SCl (streams 1-6 and 1-7 in Paper III, Table 4).}
  \label{fig:pleiades1}
  \end{center}
\end{figure}
\begin{figure}
  \begin{center}
   \epsfig{file=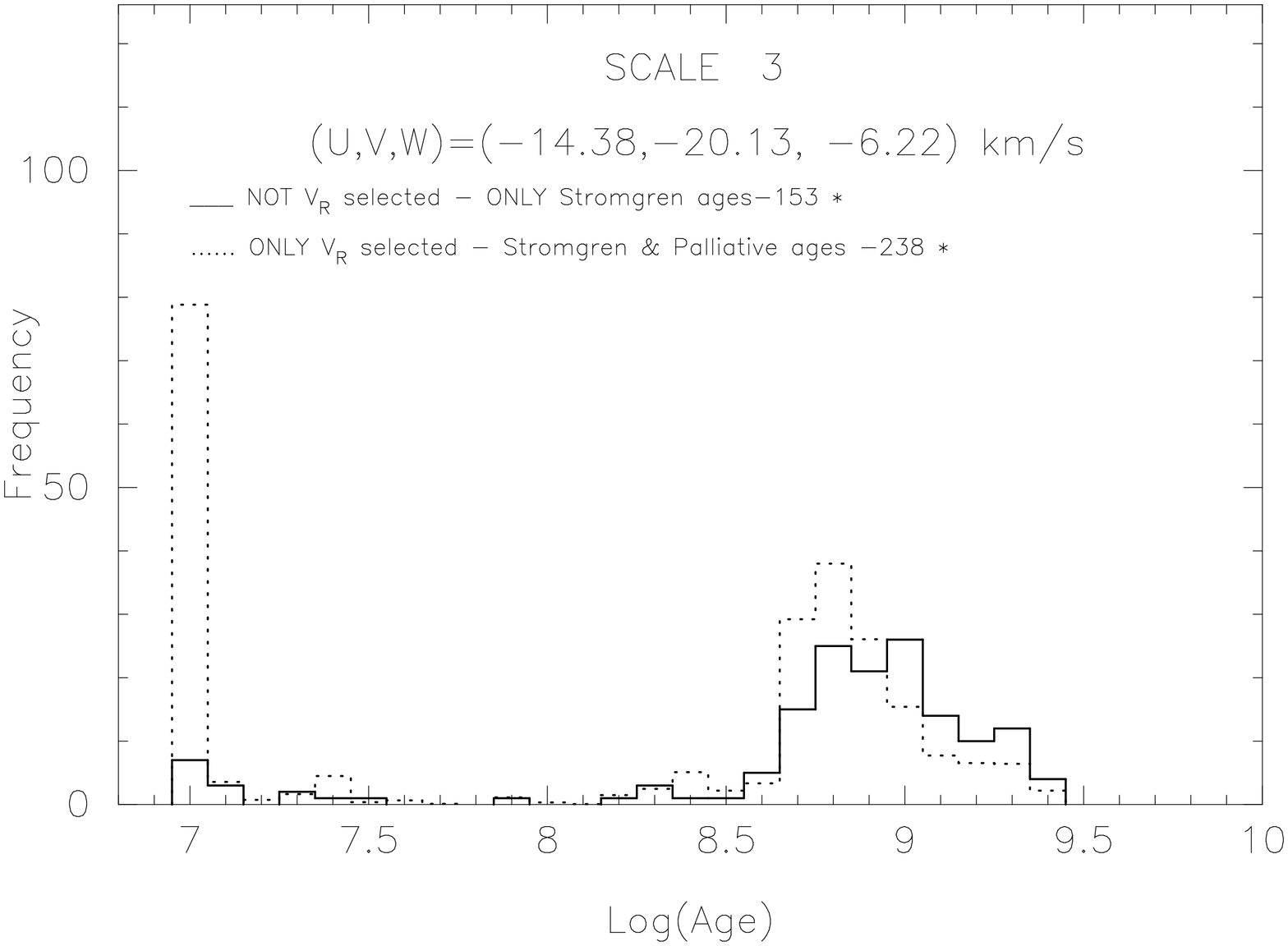,height=8.cm,width=7.cm,angle=-90.} 
  \epsfig{file=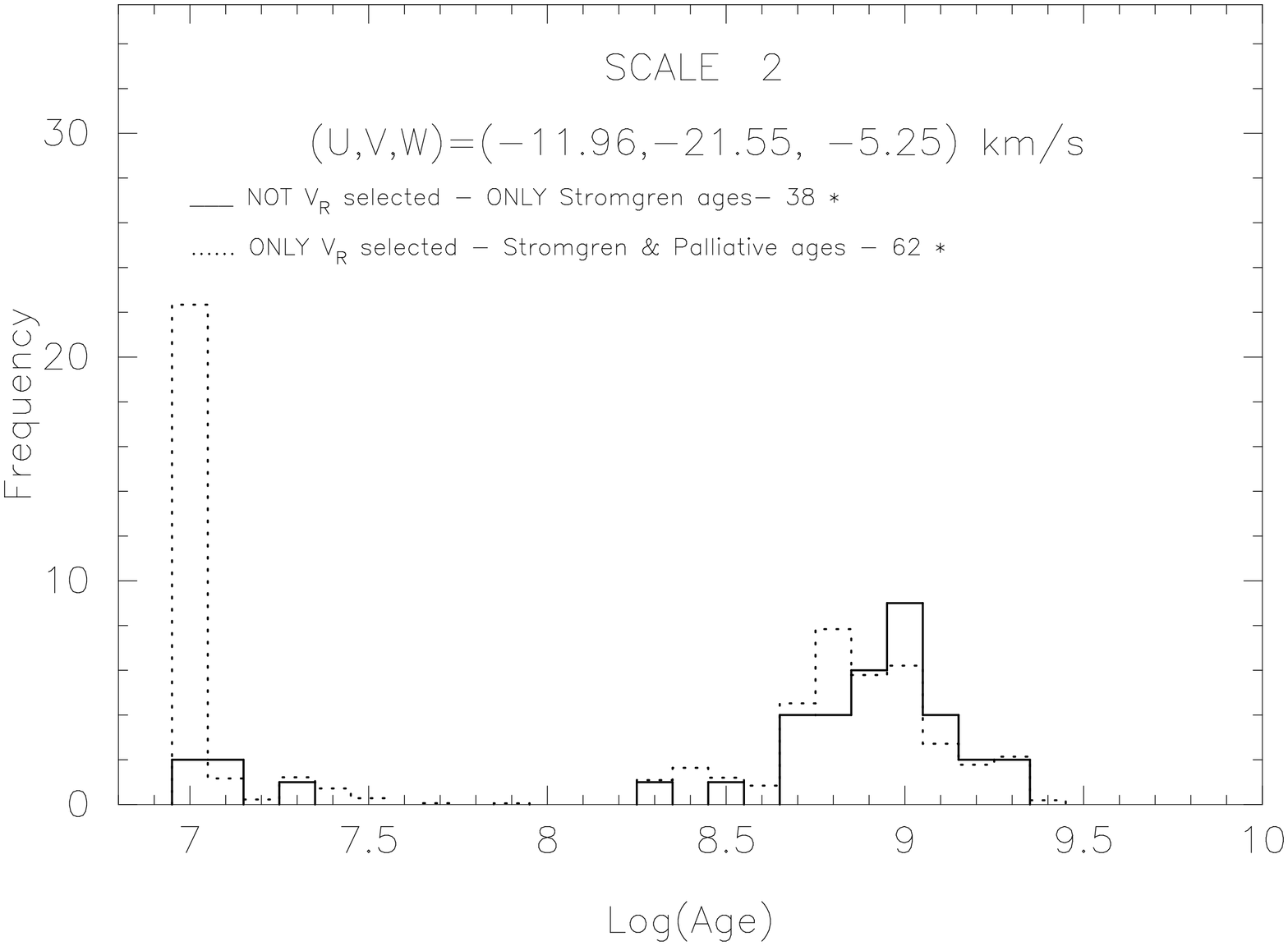,height=8.cm,width=7.cm,angle=-90.}
  \caption{\em {\bf Pleiades SCl}. Age distributions of the Pleiades SCl (stream 3-8 in Paper III, Table 2) at scale 3 ({\bf top}) and scale 2 (stream 2-5 in Paper III, Table 3) ({\bf bottom}) .}
  \label{fig:pleiades2}
  \end{center}
\end{figure}
\begin{figure}
  \begin{center}
  \epsfig{file=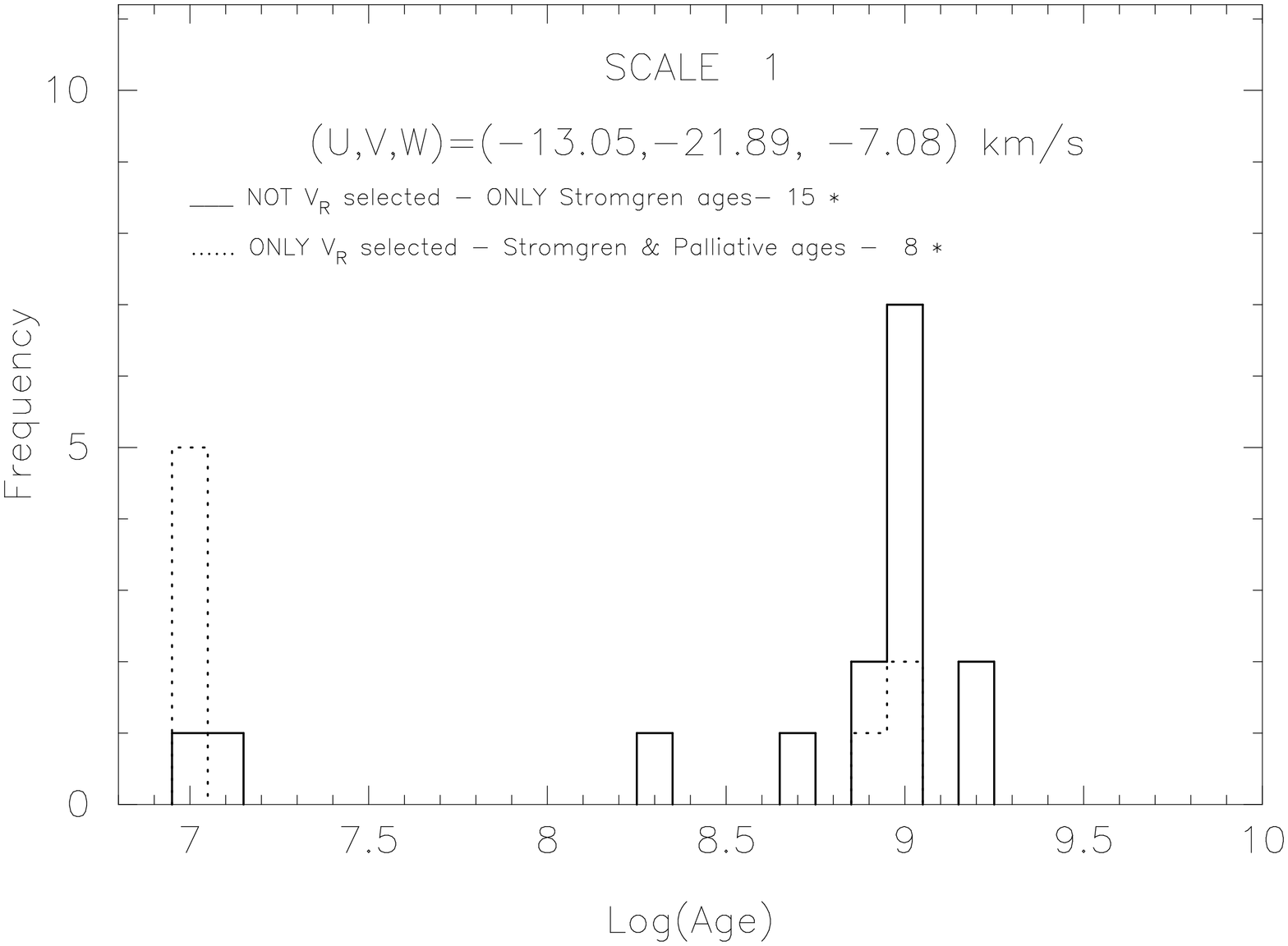,height=8.cm,width=7.cm,angle=-90.}
  \epsfig{file=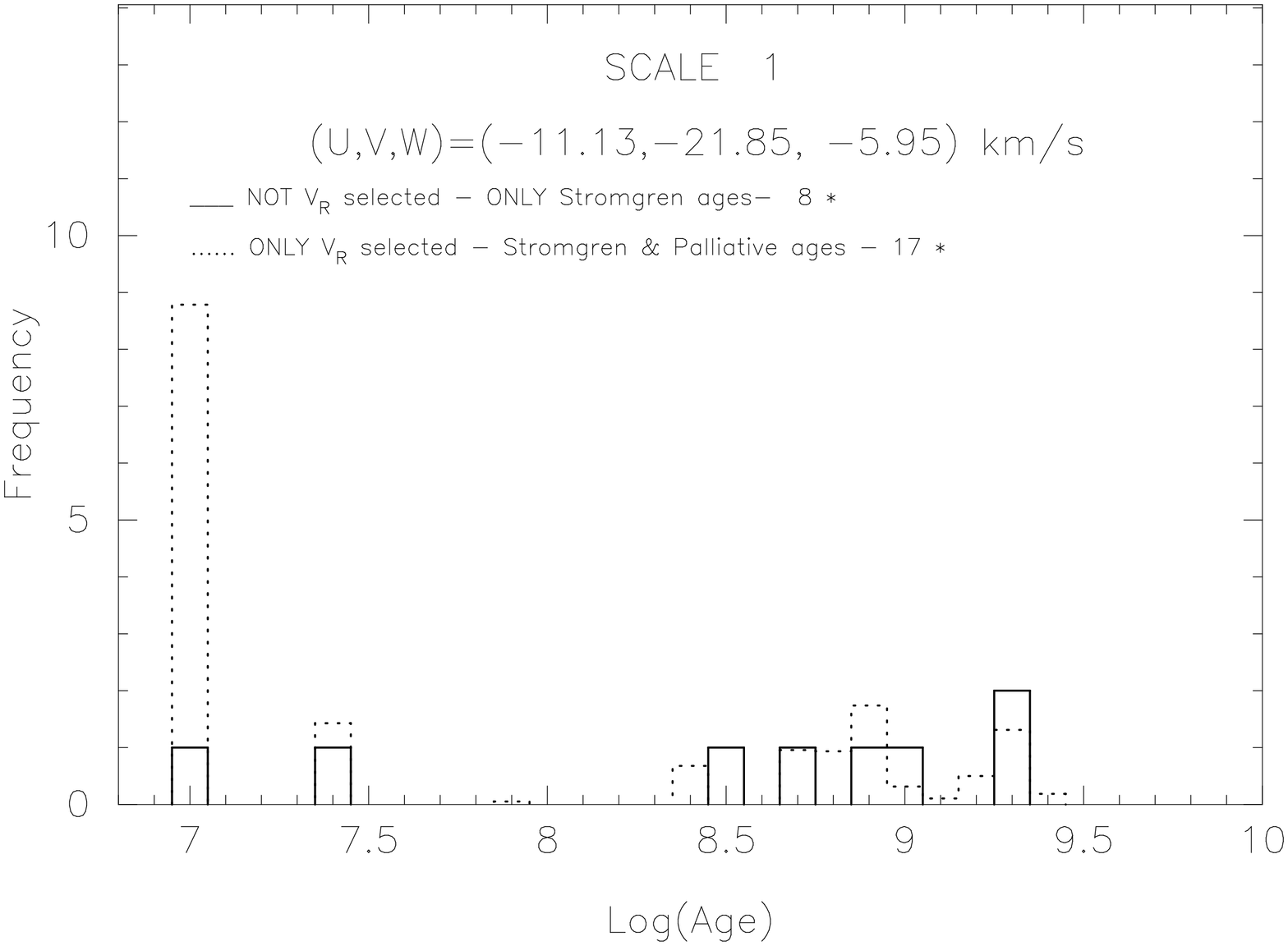,height=8.cm,width=7.cm,angle=-90.}
  \caption{\em {\bf Pleiades SCl}. Age distributions of the two sub-streams: stream 1-6 and 1-7 (in Paper III, Table 4) discovered in the Pleiades SCl at scale 1. The oldest population is mainly present in the stream 1-6 ({\bf top}) while the youngest one composes stream 1-7 ({\bf bottom}).}
  \label{fig:pleiades3}
  \end{center}
\end{figure}
\begin{figure}
  \begin{center}
\hspace*{-1.8cm}
  \epsfig{file=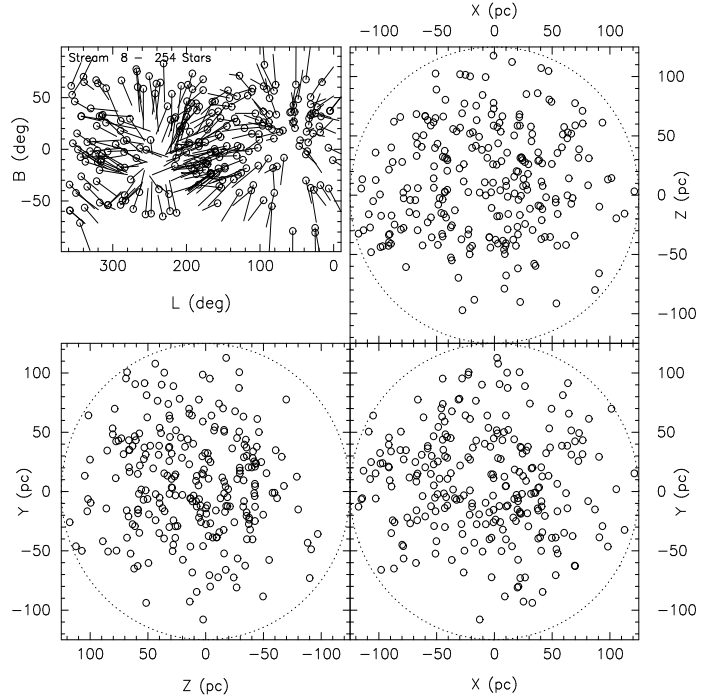,height=12.cm,width=9.4cm,angle=-90.}
  \caption{\em {\bf Space distribution of Pleiades SCl} obtained from the $V_{R}$ selected sub-sample at scale 3 (stream 3-8 in Paper III, Table 3).}
  \label{fig:spat_pleiades_s3}
  \end{center}
\end{figure}
\begin{figure}
  \begin{center}
 \hspace*{-1.8cm}
  \epsfig{file=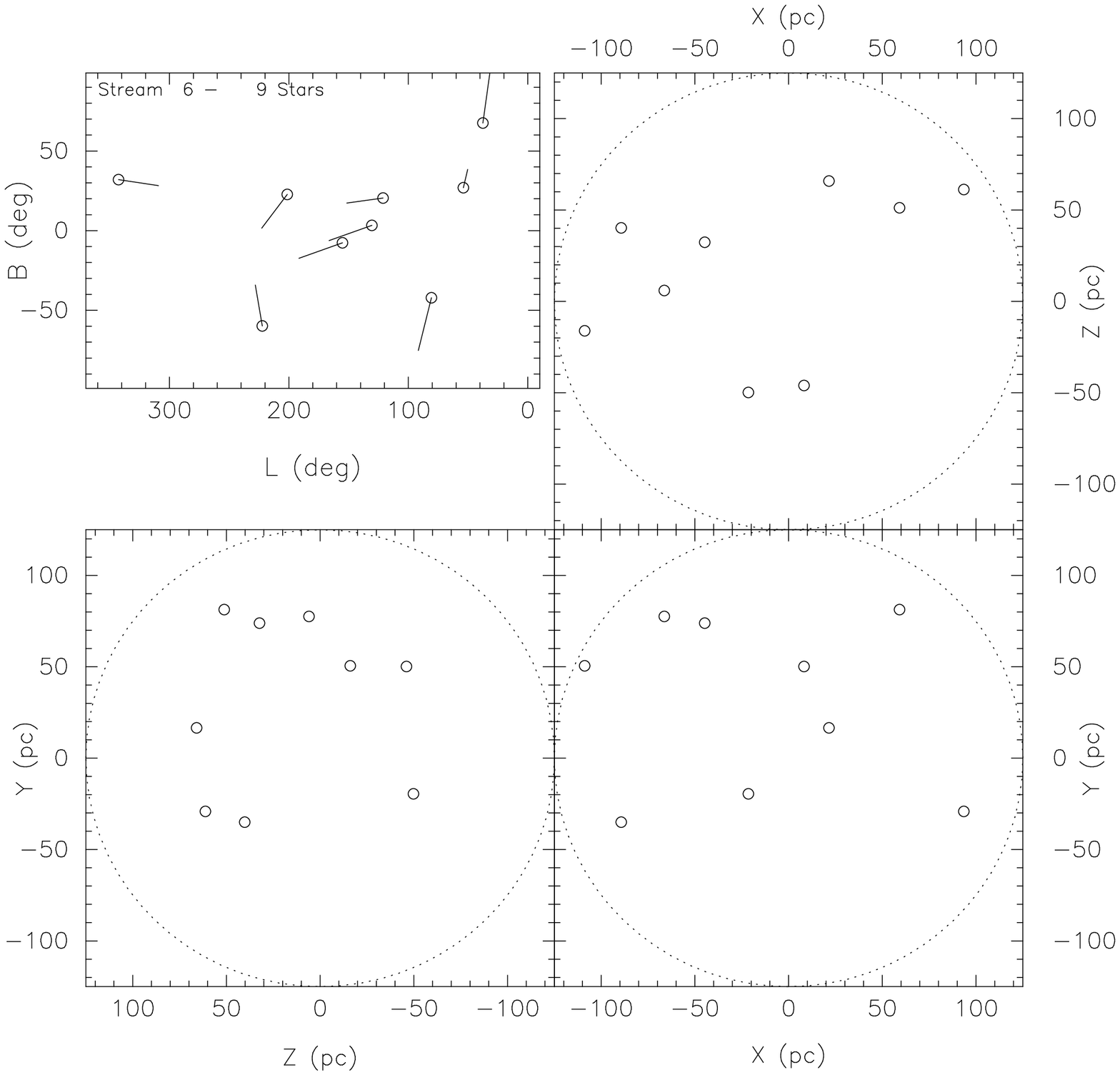,height=12.cm,width=9.4cm,angle=-90.}
 \hspace*{-1.8cm}
  \epsfig{file=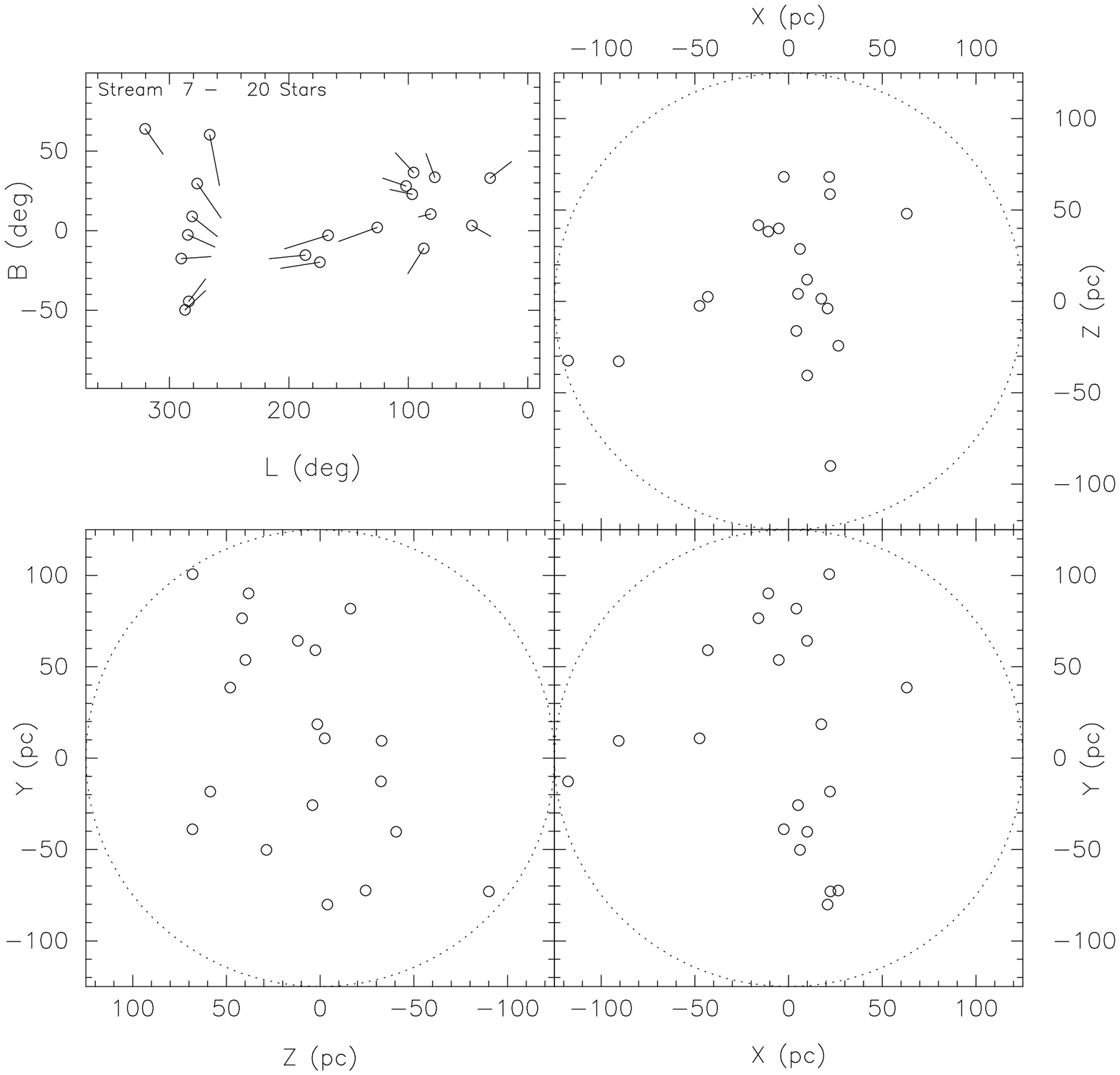,height=12.cm,width=9.4cm,angle=-90.}
  \caption{\em {\bf Space distribution of the 2 sub-streams of Pleiades SCl} from the V$_{R}$ selected sub-sample at scale 1: stream 1-6 ({\bf top}) and stream 1-7 ({\bf bottom}) in Paper III, Table 4.}
  \label{fig:spat_pleiades_s1}
  \end{center}
\end{figure}

The Pleiades {\em supercluster} (stream 3-8 in Paper III, Table 2) is detected at scale 3 
(see Figure \ref{fig:pleiades1} for velocity distributions, Figures \ref{fig:pleiades2}, \ref{fig:pleiades3} 
for age distributions and Figures \ref{fig:spat_pleiades_s3} for space distributions) and the set of stars 
selected on their observed radial component gives a mean velocity (U,V,W)=(-14.4,-20.1,-6.2) 
$km\cdot s^{-1}$ and velocity dispersions ($\sigma_{U},\sigma_{V},\sigma_{W}$)=(8.4,5.9,6.3) 
$km\cdot s^{-1}$. The age distribution (Figure \ref{fig:pleiades2}) covers the whole sample age range. 
The interval is larger than the one mentioned by Eggen (1992a): 6$\cdot 10^{6}$ to 6$\cdot 10^{8}$ yr. 
Pure Str\"omgren age distribution, peaks between $6\cdot 10^{8}$ and $10^{9}$ yr. The $V_{R}$ selected 
set with palliative ages, shows clearly the preponderance of a very young population: a peak at $10^{7}$ yr. 
A second peak at $6\cdot 10^{8}$ yr is also present. Part of this peak can be due to the statistical 
age assignation of very young stars as well as to a real stream. But this scale (scale 3) is still too 
coarse to improve previous analysis.\\
Scale 2 reveals unambiguously two groups (Figure \ref{fig:pleiades2}) of mean ages $10^{7}$ and 
$10^{9}$ yr in stream 2-5 (Paper III, Table 3). The $6\cdot 10^{8}$ year old group is no more 
present among Str\"omgren age distribution and the reminiscence found in the palliative age 
distribution is clearly due to intrinsically very young stars. At this scale, the stream is localized at 
(U,V,W)=(-12.0,-21.6,-5.3) $km\cdot s^{-1}$ with velocity dispersions 
($\sigma_{U},\sigma_{V},\sigma_{W}$)=(5.3,4.7,5.9)  $km\cdot s^{-1}$.\\ 
At the highest resolution (scale 1) the main velocity clump splits into two components. The first one, at 
(U,V,W)=(-13.1,-21.9,-7.1) $km\cdot s^{-1}$ with  ($\sigma_{U},\sigma_{V},\sigma_{W}$)=(3.1,3.3,2.5)  
$km\cdot s^{-1}$ (stream 1-6 in Table 4, Paper III) contains almost all the oldest stars ($10^{9}$ yr). 
The second component at (U,V,W)=(-11.1,-21.9,-5.9) $km\cdot s^{-1}$ with 
($\sigma_{U},\sigma_{V},\sigma_{W}$)=(1.7,3.0,1.9)  $km\cdot s^{-1}$ 
(stream 1-7 in Paper III, Table 4) is almost exclusively composed of the youngest population 
(Figure \ref{fig:pleiades3}). Moreover, the youngest component has significantly smaller velocity 
dispersions and its space distribution is more clumpy (Figure \ref{fig:spat_pleiades_s1}).\\
The interpretation comes out quite naturally. The Pleiades SCl is a chance superposition of 
two main streams originating from two different star formation epochs. One very young stream is a 
few $10^{7}$ year old and is related with the Pleiades open cluster. The relative large velocity 
difference with the Pleiades OCl ($\sim$ 8 $km\cdot s^{-1}$) and the space concentration 
(Figure \ref{fig:spat_pleiades_s1}) suggest that these stars were formed from the same interstellar 
cloud complex and at the same time as the Pleiades OCl, yet separately. The second stream is much 
older (roughly $10^{9}$ yr) and is probably related to an old cluster loosely bound by internal 
gravitation which is finishing to dissolve now. The probability to find such a coincidence in the 
velocity volume covered by the Eggen's {\em supercluster} is quite high (see Section \ref{sec:howeggenscl}).\\
\subsection{How about Eggen's superclusters}
\label{sec:howeggenscl}
The review of age and space distributions of the largest velocity structures detailed in paper III puts 
some classification in the understanding of the so-called {\em superclusters}. As it is shown in Section 
\ref{sec:pleiade_scl} for the Pleiades SCl, it is clear that most {\em superclusters}, when looked at sufficient 
resolutions (small scales), split into sub-components covering much restricted age ranges. The question then 
arises whether {\em superclusters} are real significant structures involving stars born at different epochs but 
tied to the same cell of the phase space by any binding mechanism or are they only chance coincidences between 
streams at smaller scales, each essentially coeval.  In the first case, a complex scenario such as the one proposed by 
Weidemann et al (1992) might be necessary. In such a scenario, cluster stars are formed at different epochs 
out of a single molecular cloud and remain gravitationally bound to the cloud. A process which looks hard to 
maintain over long periods.\\
It is easy to evaluate the probability that {\em superclusters} are formed by chance coincidence of smaller 
velocity dispersion streams. In Paper III, Section 5.2.2 we computed the fraction of a smooth gaussian 
3D distribution embedded in the velocity volume covered by {\em superclusters} to estimate the fraction of field stars. 
This fraction was found to be 0.192 for the 6 identified {\em superclusters}. That is an average of  
0.032 per {\em supercluster}. Under the assumption that the 38 streams appearing at scale 1 with 
$\overline{\sigma}_{stream} \sim$2.4 $km\cdot s^{-1}$ (see Table \ref{tab:table2} and Paper III, Table 4) correspond 
roughly to 38 real independent causes, most originating at different epochs 
and different places, the average number of such low dispersion streams expected to occur in a typical 
{\em supercluster} velocity volume is 38$\cdot$0.032=1.216. In Table \ref{tab:table7} we give the poissonian 
probability to get 0, 1, 2... coincidences in any {\em supercluster} velocity volume, and the number of 
{\em superclusters} built out of 1, 2,... elementary streams that one would expect to get by chance in the explored 
velocity volume. The total velocity volume contains 1/0.032=31.2 typical {\em supercluster} volumes. These average 
numbers should be compared with the observed statistics of {\em supercluster} richness in elementary streams. 
Clearly enough no meaning full test at any reasonable significance level can be built to reject the null 
hypothesis that {\em superclusters} have no physical reality.\\
  \begin{table}
 \vspace{-0.1cm}
  \caption{\em Observed number of superclusters including n or more elementary streams, compared with what would be expected if 38 independent elementary streams were distributed randomly in the whole velocity volume.}
  \label{tab:table7}
    \footnotesize  
   \begin{center}
    \begin{tabular}[h]{lcccc}
      \hline \\[-5pt]
   n &  Probability   &   Probability   & Expected number  & Observed number\\[+5pt]
      &                       &                       &   of superclusters   &  of superclusters \\  [+5pt]
      &      k=n          &     k$\geq$n   &     k$\geq$n            &       k$\geq$n \\[+5pt]
      \hline \\
 0  &0.296&1&&\\
 1&  0.360    & 0.704      &   22.0     & 26 \\
2    & 0.219   &  0.343  &  10.7 &  6\\
3     & 0.089   & 0.124   &  3.9  & 3  \\
4 & 0.027    & 0.035 & 1.1   & 1\\
5  & 0.007   &  0.008 & 0.3   &1 \\
6  & 0.001  &  0.002  &  0.1  & 0 \\ 
\hline\\
\end{tabular}
   \end{center}
   \end{table}
So {\em superclusters} most likely result from the chance coincidence in a large cell of the velocity 
space of several small streams, and the physical interpretation has to be searched for only at the 
smaller scale. Almost all the phenomena observed here can be explained by a single scenario resulting 
from two dominant mechanisms: phase mixing and cluster evaporation or disruption. At formation most stars form 
in clump of the ISM generating short lived streams which dissolve essentially over mixing time scales 
($\sim$10$^{8}$ yr): it is probably the case of the Centaurus associations and the very young 
component of the Pleiades SCl. Only streams massive enough to create some self gravitationally 
bound systems (more or less loose clusters) survive and create, as they dissolve, moderately old streams 
with age between 5$\cdot 10^{8}$ and $10^{9}$ yr (see streams in Hyades SCl, Sirius SCl and New SCl in Paper III). 
In some cases those moderately old streams can be explicitly connected to the cluster they are escaping from 
(Ursa Major OCl for the old component of Sirius SCl and possibly Coma Berenices OCl for the 
New SCl 6$\cdot 10^{8}$ year old component). However, there are a few much older groups unrelated to 
heavy {\em superclusters} (streams 3-9, 3-11 and 3-14 in Paper III, Table 2). In those rare cases, 
all connected to highly eccentric orbits, a completely different mechanism should be advocated which 
might be trapping on resonant orbits generated by the potential of the bar (Dehnen, 1998). 
%
%
\section{Conclusions}
\label{sec:conclu} 
A systematic multi-scale analysis of both the space and velocity distributions of a thin disc young star sample 
has been performed.\\
The sample is well mixed in position space since no more than 7$\%$ of the stars are in concentrated clumps 
with coherent tangential velocities. In this paper we focus on the detection of the evaporation of relatively 
massive stars (1.8 M$\odot$) out of the Hyades open cluster. The mapping we realized show an asymmetric pattern 
further forward the Hyades OCl orbit. This could be the signature of a violent encounter on one side of the open 
cluster with a massive molecular cloud. Such a picture is in agreement with the origin of streams from open cluster 
evaporation or disruption.\\
The 3D velocity field reconstructed from a statistical convergent point method exhibits a strong 
structuring at typical scales of $\overline{\sigma}_{stream}\sim$ 6.3, 3.8 and 2.4 $km\cdot s^{-1}$. 
At large scale (scale 3) the majority of structures are identified with Eggen's {\em superclusters}.
These large scale velocity structures are all characterized by a large age range which reflects 
the overall sample age distribution. Moreover, few old streams of $\sim$ 2 Gyr are also extracted at 
this scale with high U components towards the Galactic center (see Paper III). Taking into account the 
fraction of spurious members, evaluated with an observed radial velocity data set, into all these large velocity 
dispersion structures we show that they represent 63$\%$ of the sample. This percentage drops to 
46$\%$ if we remove the velocity background created by a smooth velocity ellipsoid in each structure. 
Smaller scales ($\overline{\sigma}_{stream}\sim$ 3.8 and 2.4 $km\cdot s^{-1}$) reveal that 
{\em superclusters} are always substructured by 2 or more streams which generally exhibits a 
coherent age distribution. The older the stream, the more difficult the age segregation between 
close velocity clumps inside the {\em supercluster} velocity volume. At scale 2 and 1, background stars are 
negligible and percentages of stars in streams, after evaluating the fraction of spurious members, are 38$\%$ 
and 18$\%$ respectively.\\
\\
All these features allow to describe and organize solar circle kinematics observations in a simple scenario.  
 \begin{figure}
  \begin{center}
 \hspace*{-1.cm}
  \epsfig{file=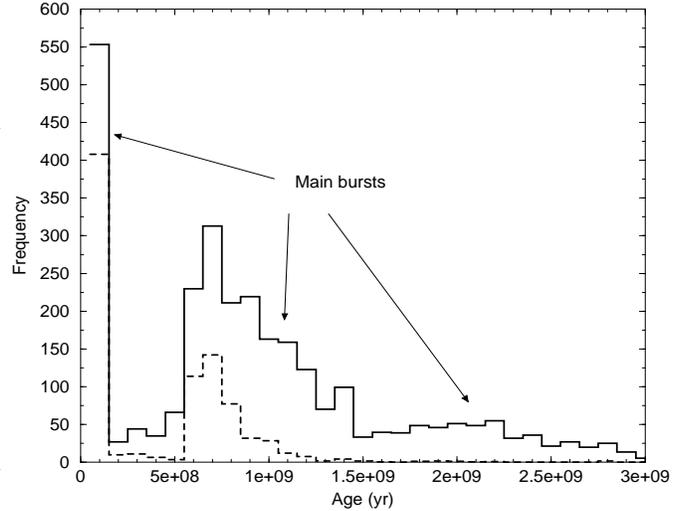,height=10.cm,width=8.cm,angle=-90.}
  \caption{\em {\bf Main bursts} in the A-F type star sample. Distribution of all the stars (full line) and stars involved in streams with typical velocity dispersion $\overline{\sigma}_{stream}\sim$ 3.8 $km\cdot s^{-1}$ (dashed line)}
  \label{fig:burst}
  \end{center}
\end{figure}  
   \begin{enumerate}
      \item Star formation in the galactic disc occurs in large bursts separated by quiescent periods 
(Figure \ref{fig:burst}). The most recent of these bursts started $\sim$ $10^{8}$ yr ago (streams in 
Pleiades SCl, IC2391 SCl and Centaurus associations) and includes the formation of the Gould Belt. 
It seems to be the start of a new active era after a half-billion year quiescent period. Within the 
look-back time of our sample, there is evidence for two long active periods, one around 1 Gyr (streams
in Hyades SCl, IC2391 SCl, Sirius SCl and New SCl) and another around 2 Gyr (the oldest streams). 
The typical duration of burst eras is also of the order of half a billion years. It is not clear, given the time 
resolution of this investigation whether during active periods the average star formation intensity oscillates 
on shorter time scales. Streams in the 1 Gyr burst reveal preferential ages at $6\cdot 10^{8}$, $8\cdot 10^{8}$ 
and $10^{9}$ yr favouring this idea of oscillations, possibly related to spiral waves. It may also reflect more 
local phenomena.\\
\item Under burst conditions, stars mainly form in groups reflecting the clumpy structure of the interstellar 
medium sporadically sampled by the star formation process. As a consequence, the velocity space is 
gradually filled by successive star formation bursts. Formation puts stars preferentially on near circular orbits
filling the center of the velocity ellipsoid (young streams in Pleiades SCl, IC2391SCl and Centaurus). 
About 75$\%$ of recently formed stars belong to streams which internal velocity dispersions do 
not exceed 4 $km\cdot s^{-1}$. A limited fraction of the initial groups are gravitationally bound and
form open clusters.\\
\item Open clusters sustain a stream of stars with similar velocity by an evaporation process due to internal 
processes or tidal disruption by gravitational potential large scale inhomogeneities (encounters with GMCs). 
This phenomenon is well illustrated by the needlelike density structure found around the Hyades OCl. Streams found at 
scale 1 and 2 ($\overline{\sigma}_{stream}\sim$ 2.3 and 3.8 $km\cdot s^{-1}$) seem to be in 
agreement with this scenario despite the fact that age and membership accuracy does not always 
permit to show clear correlation between age and velocity.\\
\item In this picture the survival of streams as old as $10^{9}$ yr is satisfactorily explained as the 
end of the evaporation process of the most concentrated clusters. Observational evidences obtained by 
\cite{Wielen71} and \cite{Lyn82} show that the half lifetime of open clusters in the solar neighbourhood is 
between 1-2 10$^{8}$ yr.  A longer survival for streams can be explained in terms of resonant trapped orbits by the 
Galactic bar gravitational potential. Indeed, all the streams 
older than $10^{9}$ yr are on the external part of the velocity ellipsoid with a high absolute value of 
their U component. They probably have been formed in the inner part of the disc where the bar potential 
is capable to lock them into an orbital resonance (Dehnen, 1998).\\
\item The typical scale of Eggen's {\em superclusters} 
($\overline{\sigma}_{stream}\sim$ 6.3 $km\cdot s^{-1}$) does not seem to correspond to any physical 
entity. For one thing, the picture they form, their frequency and their divisions at smaller scales are 
compatible with their creation by chance coincidence of physically homogeneous smaller 
scale structures  ($\overline{\sigma}_{stream}\sim$ 3.8 or 2.4 $km\cdot s^{-1}$). For the other thing, 
their internal age distributions more or less reflect the overall age distribution of the whole sample, 
with occasionally some preference for the typical age of the dominant sub-structure.\\
\end{enumerate}
Beyond this phenomenological classification, the 6D analysis of this complete sample of nearby A-F 
stars provides the first time dependent picture of the mechanism creating the stellar velocity 
distribution in the disc. 
%
%
\begin{acknowledgements}
This work was facilitated by the use of the Vizier service and the Simbad database developed at CDS.
E.C thanks A. Bijaoui and J.L. Starck for constructive discussions on wavelet analysis, and C. Pichon 
for interesting discussions and help on some technical points. E.C is very grateful to R. Asiain 
for providing the two codes of age calculation from Str\"omgren photometry and thanks 
J. Torra and F. Figueras for valuable discussions.
\end{acknowledgements}

\end{document}